\documentclass[twocolumn,10pt, reprint, nofootinbib, amsmath, amssymb, aps]{revtex4-2}
\usepackage{comment}
\usepackage{amssymb}
\usepackage{amsmath}
\usepackage{bm}
\usepackage[dvipsnames]{xcolor}
\usepackage{soul}
\usepackage[linkcolor = blue, citecolor = blue, urlcolor = blue, colorlinks = true]{hyperref}
\usepackage{graphicx}
\usepackage{etoolbox}
\usepackage{tikz}
\newrobustcmd*{\mysquare}[1]{\tikz{\filldraw[draw=#1,fill=#1] (0,0)
rectangle (0.2cm,0.2cm);}}

\begin{document}

\title{Phase Behaviour and Dynamics of Three-Dimensional Active Dumbbell Systems}

\author{C. B. Caporusso$^{1,\dag}$, 
G. Negro$^{1,\dag}$\footnotetext[0]{These authors contributed equally.},
A. Suma$^{1}$,
P. Digregorio,
L.~N. Carenza$^{2}$,
G. Gonnella$^1$,
L.~F. Cugliandolo$^{3,4}$ 
}
\affiliation{
 $^1$Dipartimento di Fisica, Universit\'a degli Studi di Bari and INFN, Sezione di Bari, via Amendola 173, Bari, I-70126, Italy, \\
 $^2$Instituut-Lorentz, Universiteit Leiden, P.O. Box 9506, 2300 RA Leiden, Netherlands,\\
 $^3$Sorbonne Universit\'e, Laboratoire de Physique Th\'eorique et Hautes Energies, CNRS UMR 7589, 4 Place Jussieu, 75252 Paris Cedex 05, France
 \\$^4$Institut Universitaire de France, 
  1 rue Descartes, 75231 Paris Cedex 05, France
}
\begin{abstract}
We present a comprehensive numerical study of the phase behavior and dynamics of a three-dimensional active dumbbell system with attractive interactions. We demonstrate that attraction is essential for the system to exhibit nontrivial phases.
We construct a detailed phase diagram by exploring the effects of the system's activity, density, and attraction strength. We identify several distinct phases, including a disordered, a gel, and a completely phase-separated phase. Additionally, we discover a novel dynamical phase, that we name percolating network, which is characterized by the presence of a spanning network of connected dumbbells. In the phase-separated phase we characterize numerically 
and describe analytically the helical motion of the dense cluster. 
\end{abstract}

\maketitle


\section{Introduction}
\label{sec:introduction}


Active systems are a striking class of soft matter which employs internal energy stored in some kind of reservoir and converts it into motion~\cite{marchetti2013,Bechinger2016,Shankar2022,carenza2019}. 
Active systems greatly vary in their extension, ranging from macroscopic to microscopic assemblies of active constituents, and a multitude of different realizations. Focusing on the microscopic world, active systems can either have synthetic origin (\emph{e.g.} Janus particles~\cite{Walther2013,Das2015,ebbens2016}) or biological origin. In this latter case, cytoskeletal suspensions~\cite{sanchez2012} and cellular cultures~\cite{bookhoward,saw2017,Guillamat2022,Armengol2023} are examples of \emph{in vitro} systems which are often employed to study active behaviors in a controlled environment, but more complex living organisms, such as developing embryos  or living tissues~\cite{Haeger2019,MaroudasSacks2021,Brauns2023} have recently been considered for their promising implications in understanding a variety of different biological phenomena from morphogenesis to cancer progression and spreading of infections in living organisms.
Regardless of their particular realization, active systems can exploit energy to interact with the surrounding environment and perform autonomous motion, leading to a plethora of collective behaviors.


In the past two decades much effort has been spent to identify the physical rules underlying the behavior of living and active systems~\cite{marchetti2013,Hatwalne2004,Kruse2004,cates2015}, with important repercussions on our understanding of their non-equilibrium dynamics. Among others, bacteria have gained the attention of the physics community as they represent a simple natural realization of self-propelled particles, by virtue of their biological simplicity and limited ability in interacting with both the environment and other units. Therefore, bacterial systems provide an elementary yet relevant example of how the dynamics of active constituents may lead to self-assembly~\cite{Bratanov2015,Bi2018,MassanaCid2022}. 
Indeed bacteria exhibit  {noteworthy} chemotactic properties which allow them to respond to external stimuli such as variations of temperature, nutrient availability, oxygen concentration, \emph{etc.} to develop different kinds of colonies, ranging from biofilms to fluidic suspensions, often characterized by the emergence of elaborate patterns, even in absence of guidance and only because of the uncoordinated evolution of the separate units~\cite{Yaman2019}. 

The chemotactic properties of bacteria are often exploited in experiments to trigger a particular response. For instance, most bacteria preferentially reproduce at room temperature --process during which they mostly remain motionless~\cite{Basaran2022a,Basaran2022b}-- while at higher temperatures they can either develop biofilms if the growing substrate is dry or become motile by growing flagella --a proteic protrusion which is used by some bacterial species to swarm and swim in  {wet} environments. In this latter case, oxygen availability or light intensity determines the dynamical response of the system and can be used in experiments as a control parameter to tune the typical speed of migration. When this is large enough, dense suspensions of flagellate bacteria in a fluidic environment arrange in a liquid crystalline fashion~\cite{Yaman2019} and develop a chaotic flowing state characterized by whirling patterns which resemble those observed in an isotropic fluid flowing at large Reynolds numbers --a feature that earned this dynamical state the title of bacterial (or active) turbulence~\cite{Dunkel2013,giomi2015,carenza2020,carenza2020_bif,Aranson2022,Alert2022}.



Inspired by this huge variety of possible states, many models have been advanced to capture the dynamics of bacterial suspensions. They range from particle (or molecular) models~\cite{marchetti2013,Locatelli2015,Solon2015} --where each unit in the system evolves according to a given equation of motion and interactions-- to continuous models~\cite{cates2022,wittkowski2014,Arciprete2018,negro2019b,Giordano2021,Carenza2020b,Favuzzi2021,Carenza2019d} --where the status of the system is described in terms of a few continuous fields capturing only the slowly-varying hydrodynamic features of the system (local group velocity, polarization, density, \emph{etc.}). 

In the following we will take the former \emph{molecular} approach and describe every constituent as a dumbbell --namely two joint spheres linked with a rigid spring -- autonomously moving in a viscous medium and subject to thermal fluctuations. This particular realization 
falls in the wide class of the so-called \emph{self-propelled active Brownian} constituents. 
Many recent studies have focused on different elements ranging from disks~\cite{digregorio2019,caporusso2020,Caporusso-2023,cagnetta2017,cates2015} to rods and ellipsoids~\cite{Bar2020,speckPolarPRE20}, and contributed to explicate the effects of non-equilibrium in phenomena like motility induced phase separation (MIPS)~\cite{cates2015} and arrested phase separation~\cite{Sciortino2008b}, providing significant insights on the physics of collective effects of bacterial suspensions and other similar active systems. Interestingly enough, it was repeatedly suggested that active systems could hinge on a minimal toolkit of physical mechanisms, independently of the particular realization, to achieve a specific dynamical response.
However, most research has focused on two-dimensional systems, while much less is known on three-dimensional realizations, relevant for a full characterization of real biological systems, as these are most often subject to a full $3D$ dynamics.

To better outline the scope of our study, we find it useful to summarize some important results obtained in systems of active Brownian particles (ABP) both in $2D$ and $3D$, before proceeding with the presentation of our findings. In Sec.~\ref{subsec:summary-old}, we provide a brief summary of previous studies, regarding the properties of MIPS at varying the system's dimensionality and the particles' interaction.  Far from being a complete review on ABP, we suggest the interested reader to refer to Ref.~\cite{cates2015}. Next, in Sec.~\ref{subsec:summary-new} we motivate our 
study and summarize our results.
 
 \subsection{Brief summary of previous studies}
 \label{subsec:summary-old}
 


Most literature on active Brownian systems focused on two-dimensional realizations, both for their biological relevance and for the important implications on the melting mechanism in $2D$~\cite{Halperin78}. 
In recent years, there has been significant progress in understanding the non-equilibrium behavior of active Brownian particles, with much attention on disks and dumbbells. Both types of systems exhibit 
 {segregation into dense and dilute phases in absence of attractive interactions, solely due to persistence in motion.}
This segregation phenomenon, known in the literature as motility-induced phase separation (or shortly MIPS), consists of 
 {the interplay between two processes.}
First, scattering of active particles results into 
 {a local drop of velocity due to mutual repulsion}; second, 
particles tend to aggregate into dense clusters 
 {due to their persistence.}
A positive feedback  {between these two processes drives nucleation and growth of clusters and ultimately a macroscopic phase separation. Clusters in turn}
displace and also aggregate~\cite{caporusso2022dynamics, Caporusso-2023}.
More specifically, for dumbbells a coexistence region between dense (hexatic) and dilute (liquid) phases, found in the absence of activity, extends continuously to high values of activity~\cite{cugliandolo2017}. For disks, instead, between the hexatic-liquid coexistence and MIPS, occurring at low and high activity, respectively, there is a region at intermediate activity where no coexistence between different phases is found~\cite{digregorio2018}.
Other special features of the active disk systems include micro-separation of hexatic domains or the formation of cavitation bubbles in the interior of the dense aggregates~\cite{digregorio2018,cugliandolo2017,caporusso2020} which take different form in active diatomic systems~\cite{sum_gonn_mar, Suma2014, Suma2014pre, petrelli2018, Caporusso-2023}.

While the panorama is pretty much understood for $2D$ realizations of active systems, much less is known about the phase behaviour of these same systems in $3D$~\cite{Stenhammar2014, wysocki2014, vandamme2019, nie2020, omar2021,Turci21,Venkatareddy23}.  Stenhammar \emph{et al.}~\cite{Stenhammar2014} first observed that the MIPS transition is significantly  {inhibited} in $3D$ for active Brownian  {spheres} in absence of any attractive interactions
 {and shifted towards higher activities, due to reduced persistence in $3D$ space}.
 {On the other hand}, van Damme \emph{et al.}~\cite{vandamme2019} pointed out that MIPS is   {completely} suppressed 
for spherocylinders  {of aspect ratio $2$} in $3D$.
The authors argued that MIPS suppression is ultimately due to 
passive torques  {originating from anisotropic steric repulsion between the rods.}
 {This makes rods quickly reorient upon binary collisions or encounters with a cluster, preventing accumulation and phase separation.}

Three-dimensional systems of active spherical particles~\cite{Mognetti2013} and dumbbells~\cite{schwarz2012} in presence of attractive interactions have been studied numerically with the purpose of addressing the experimental realization of bacterial systems in interaction with non-absorbing polymers~\cite{Melaugh2019,schwarz2012}. 
In this case, demixing into coexisting phases is tamed by the aggregation strength, leading to the formation of highly dynamical structures as rotating clusters~\cite{schwarz2012}. However, activity tends to suppress aggregation, consistently with experimental observations of suspensions of motile bacteria. The non-equilibrium nature of the phase coexistence is also illustrated by the formation of new dynamical states, such as percolating networks 
caused by the interplay between attraction and motility~\cite{prymidis2015}.


Furthermore, the phase behavior of $3D$ passive attractive atomic and molecular systems is very rich in itself and is not fully understood yet~\cite{Zaccarelli2007}. 
This is because the way in which many-body systems interact is highly influenced by the geometry of the elementary constituents and different macroscopic behaviors are observed in systems with different shapes.  
The interest in these systems 
 {has grown over} the last 15 years or so, thanks to the possibility of engineering colloids with different shapes, sphere's diameter and separation length, and tuning particle-particle interactions~\cite{Kraft12}, paving the way towards the design of colloidal crystals with useful optical properties.
Indeed, dumbbells model diatomic molecules and show interesting crystal structures at equilibrium~\cite{Vega92,Marechal08,Kowalik08,Dennison12}.
In the context of self-assembly studies~\cite{Jack15}, patchy dumbbells have been the focus of much attention~\cite{Whitelam10, Munao13, Munao14, Munao15, Avvisati15, OToole17}. 
In the limit of weak activity, the $3D$ model that we will consider in this article is relevant to the description and characterization of these very interesting passive systems.
 
\subsection{ {Goal and article structure}}
\label{subsec:summary-new}
 

In this introduction, we discussed how clustering of bacterial suspensions can be explained in terms of simple physical mechanisms
when modelled as systems of active Brownian constituents. 
Importantly, most of these results and observations only hold in two-dimensions and cannot be invoked to explain 
the coarsening of bacterial aggregates
in three-dimensions.
However, collective effects in bacterial suspensions are also observed in $3D$. 



Our goal in this paper is to clarify the relevance of motility in the dynamics of three-dimensional suspensions by performing a systematic analysis of the phase behavior of a system of propelled attractive dumbbells. 

The article is structured as follows. 
In Section~\ref{sec:model} we will present the model for active Brownian dumbbells.
The discrete mesoscopic approach here implemented allows us to retain only the main features of bacterial suspension (local orientation, attractive interactions) 
and to wipe out more system-dependent features. Section~\ref{sec:bulk}
is devoted to the analysis of the phase behavior of the system in a three-dimensional unconfined geometry. 
We will identify four phases (gel, disordered, phase-separated and 
percolating network) while varying two control parameters, namely the P{\' e}clet number ${\rm Pe}$ measuring 
 {the strength of the activity} with respect to the thermal fluctuations and the total packing fraction $\phi$. The effect of the attraction strength will also be considered. 
Next, in Sec.~\ref{sec:single-cluster} we will focus on the description of the motion of a single isolated cluster in the phase-separated phase. 
We close the paper with some conclusions in Sec.~\ref{sec:conclusions}.







\section{Model and numerical methods}
\label{sec:model}

\subsection{ {Model}}

We simulate a system of $N$ dumbbells immersed in a three dimensional space. 
 {Each dumbbell is a dimer composed of two spherical beads separated by a  distance $\sigma$, and each of mass $m$.} The beads are rigidly connected and their center-to-center distance is kept fixed at distance $\sigma$. Each bead $l$ evolves in time according to the Langevin equation 
\begin{equation}
    m\ddot{\bm{r}}_l = -\gamma \dot{\bm{r}_l} + {\bm{f}_{\text{act}}}_l - \bm{\nabla}_l U + \bm{\xi}_l
    \; , 
    \label{eq:langevin}
\end{equation}
where $l=1,..,2N$. $\gamma$ is the damping coefficient and $U$ is the total potential energy. The term $\bm{\xi}_l$ is a Gaussian white noise, with zero average and variance fixed by the fluctuation-dissipation theorem as 
\begin{eqnarray}
    \langle \xi_{l\alpha}(t) \rangle &=& 0 
    \; , \\
    \langle \xi_{l\alpha}(t_1) \xi_{p\beta}(t_2) \rangle &=& 2k_B T \gamma \delta_{lp}\delta_{\alpha\beta}\delta(t_1-t_2)
    \; ,
\end{eqnarray}
where $\alpha,\beta=1,2,3$ label the spatial coordinates, $T$ is the temperature of the equilibrium bath and $k_B$ the Boltzmann constant.  

\begin{figure}[t!]
  \centering
  \includegraphics[width=0.78\columnwidth]{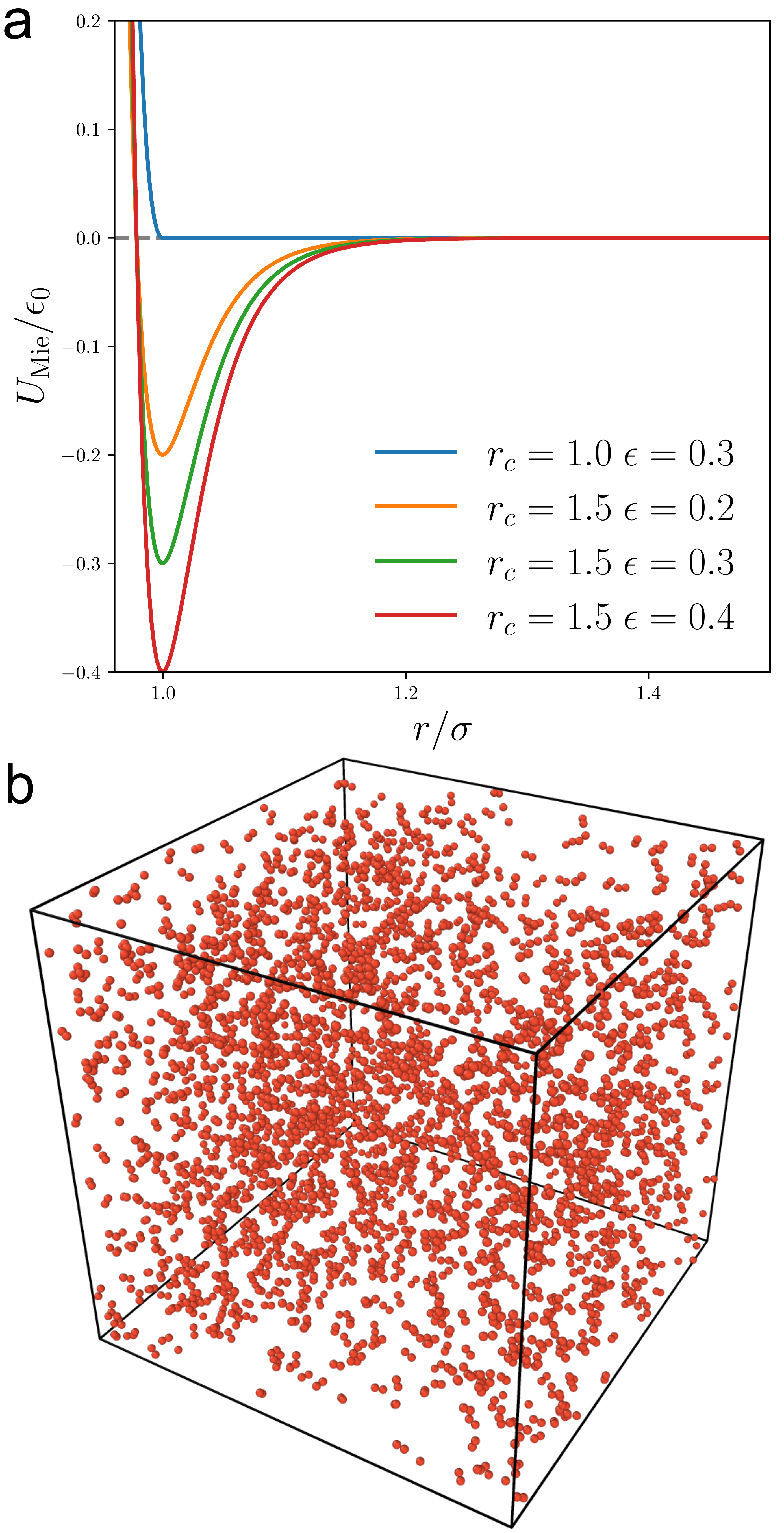}
  \caption{{\bf Simulation set up}. 
   { Mie potential shown for four combinations of the cutoff length, $r_c$, and of the attraction strength, $\epsilon$. These values are reported in the key in units of length, $\sigma$, and of energy, $\epsilon_0$. All curves are shifted so that $U_{\text{Mie}}(r_c)=0$.  In all cases $n=32$. }
   (b) A snapshot of an active dumbbell system with pure repulsive interbead interactions in a $3D$ periodic box. 
  $\phi=0.01$ and Pe $=150$. Without attraction between the dumbbells MIPS is absent even at higher densities and activity.   
}\label{fig:1}
\end{figure}

The self-propulsion is represented by the active force ${\bm{f}_{\text{act}}}_l$. This force has the same fixed modulus $f_{\rm act}$ for all  {spheres}. It points from the tail to the head monomer in the dumbbell and acts along the line that joins the centers of the two beads. 
The tail and head identity of the 
beads in a molecule are attributed randomly at the beginning 
of the simulation and they are kept fixed along the numerical experiment.

Beads belonging to different dumbbells interact through the Mie short-range potential:
\begin{equation} 
\begin{split}
 {U_{\text{Mie}}(r)} &= 4\epsilon  {\epsilon_0}\left[ \left(\frac{\sigma_{\text{Mie}}}{r}\right)^{2n} - \left(\frac{\sigma_{\text{Mie}}}{r}\right)^{n} \right]\theta(r_c-r) \\
 & {-4\epsilon  {\epsilon_0}\left[ \left(\frac{\sigma_{\text{Mie}}}{r_c}\right)^{2n} - \left(\frac{\sigma_{\text{Mie}}}{r_c}\right)^{n} \right]}
 \; , 
\label{Mie}
\end{split}
\end{equation}
 {with $n=32$, $\sigma_{\text{Mie}}=2^{-1/n}\sigma$, $r$ the distance between the concerned beads, $r_c$ the potential cutoff, $\theta(r)$ the Heaviside step function and $\epsilon_0$ the energy unit. The second term on the right-hand side represents  a shift in the potential such that $U_{\text{Mie}}(r_c)=0$. 
The time unit is defined as $\tau = \sqrt{m\sigma^2/\epsilon_0}$.}
The variation of the potential $U_{\rm Mie}$ 
with the parameter 
$\epsilon$ controlling the potential energy scale is shown in Fig.~\ref{fig:1}(a). 
This potential has a pronounced minimum 
at $r_{\rm min}= 2^{1/n} \sigma_{\rm Mie} = \sigma$ and it allows colloids to behave similarly to hard  {spheres} at distances shorter than $\sigma$~\cite{digregorio2018,cugliandolo2017}, while it acts attractively at longer distances. Unless otherwise stated, we use $r_c=1.5\sigma$. 

The system can be characterized through the following dimensionless numbers.
The relevance of active injection with respect to thermal fluctuations is captured by the P\'eclet number
\begin{equation}
    \text{Pe} = \frac{2f_{\text{act}} \sigma}{k_B T}
    \; , 
\end{equation}
which is the dimensionless ratio between the advective transport rate and the diffusive transport rate  {(the constant two relates to the total active force on the dumbbell, which is twice $f_{\text{act}}$)}. Moreover, the aggregation capability of the system can be captured by measuring the ratio between the strength of the attractive force and the active force~\cite{Mognetti2013} 
 \begin{equation}
     P_{\rm agg}=\frac{\epsilon}{f_{\rm act}\sigma} 
     \; . 
 \end{equation}
Finally, the global density or packing fraction  is quantified by the volume fraction
\begin{equation}
    \phi = N \, \frac{\pi \sigma^3}{3V} \; ,
\end{equation}
 with $V$ the volume of the box where the dumbbells are placed
 and $N$ the total number of dumbbells, that is, twice the number of spherical beads.  {Although in the following we will mostly consider relatively low packing fractions, $\phi \leq 0.45$, it is worth mentioning that the close-packing value $\phi_{\rm cp} \simeq 0.74$ and the random close packing value $\phi_{\rm rcp} \simeq 0.64$ for hard spheres in $3D$. Close packing is achieved by the face center cubic and the hexagonal close packing arrangements. In both cases each sphere touches 12 neighboring spheres~\cite{Paddy23}.}

  {From now on we set the units of length, energy and mass, respectively $\sigma$, $\epsilon_0$ and $m$, to 1.}

\subsubsection{ {Numerical details}}

 {The evolution equations are integrated using the velocity-Verlet algorithm with the open-source code LAMMPS~\cite{LAMMPS}. Both beads of each dumbbell are constrained at fixed distance via the SHAKE algorithm, which applies at each timestep an additional force so that the bond length remains constant at the next iteration step~\cite{refSHAKE}. The integration timestep is $\delta t = 0.005$. The dumbbells move in a cubic box with fully periodic boundary conditions.} 

 {We use as attraction strengths $\epsilon=0.3$, though we have also explored cases with $\epsilon=0.2$ and $\epsilon=0.4$. We set $k_BT=0.05$ and $\gamma=10.0$. The latter choice ensures that, although we are considering the inertial contribution in Eq.~(\ref{eq:langevin}), the dynamics is over-damped over relatively short timescales of the order of $m/\gamma\sim 0.1$.}

 {Simulations are performed fixing $N=4096$ and varying the box size to match the target volume fraction $\phi$. Densities are varied between $0.001$ to $0.45$, while the P\'eclet number Pe is varied between $0$ and $200$. 
Simulations are started from dumbbells placed randomly in the box, and evolve until stationary conditions are reached. This typically occurs after  $10^5$ time units. Afterwards, systems are simulated typically for $5\cdot10^5$ time units in order to collect data. Stationary conditions are checked by looking that the probability density distribution does not vary looking at different times during simulation. Note that in the case of the gel phase described in Section~\ref{sec:bulk} 
the system is in an arrested state. 
While we observe slight changes in the potential energy as a function of time, due to small rearrangements of particles inside the gel branches,  the density distribution is substantially stationary. We do however observe that the thickening of branches is slightly faster increasing Pe.}


We apply the DBSCAN algorithm for cluster identification \cite{dbscan, digregorio2019, caporusso2020}. DBSCAN is a density-based clustering algorithm that can identify clusters of arbitrary shape and size, grouping together points that are closely packed 
and marking as outliers points that lie alone in low-density regions. 
DBSCAN works with two parameters: $R$ and $n_{\text{min}}$. $R$ specifies the radius of a neighborhood around a point, and $n_{\text{min}}$ specifies the minimum number of points required to form a dense region. A point is classified as a {\it core point} if there are at least $n_{\text{min}}$ other points within $R$ distance from it. A point is classified as a {\it border point} if there are less than $n_{\text{min}}$ points within $R$ distance from it, but is within $R$ distance of a core point. A point is classified as a {\it noise point} if it is neither a core nor a border point.
In our case, we choose $R=1.5\sigma$ 
and $n_{\text{min}}=12$, based on the preferred structures at close packing. 
Then, DBSCAN assigns each core point to a cluster, along with all the points that are density-reachable from it (i.e., there is a path of core points connecting them). Border points are assigned to the cluster of their nearest core point, and noise points are left unassigned.  {Note that changes in the parameters can affect the identification of clusters. For instance, increasing the radius to values larger than $2$, or decreasing $n$  leads to the incorrect identification of neighbouring clusters as the same one. Instead, decreasing the radius below $1$ or increasing the required number of neighbours inhibit the identification of the clusters. We visually verified that the assigned clusters are identified correctly with this choice of parameters.}


\section{Phase Behaviour}
\label{sec:bulk}

 {Before entering into the heart of our study, we confirmed that there is no stable MIPS~\cite{Turci21} for purely repulsive 
active dumbbells.
The condition for purely repulsive dumbbells is achieved by setting the cut-off lengthscale $r_c=1$, so that only the repulsive core in the Mie potential is preserved (blue curve in Fig.~\ref{fig:1}(a)). A typical configuration of the steady state is shown in Fig.~\ref{fig:1}(b), clearly demonstrating the absence of particle  aggregation}. 

\begin{figure*}[t!]
  \centering
  \includegraphics[width=2.0\columnwidth]{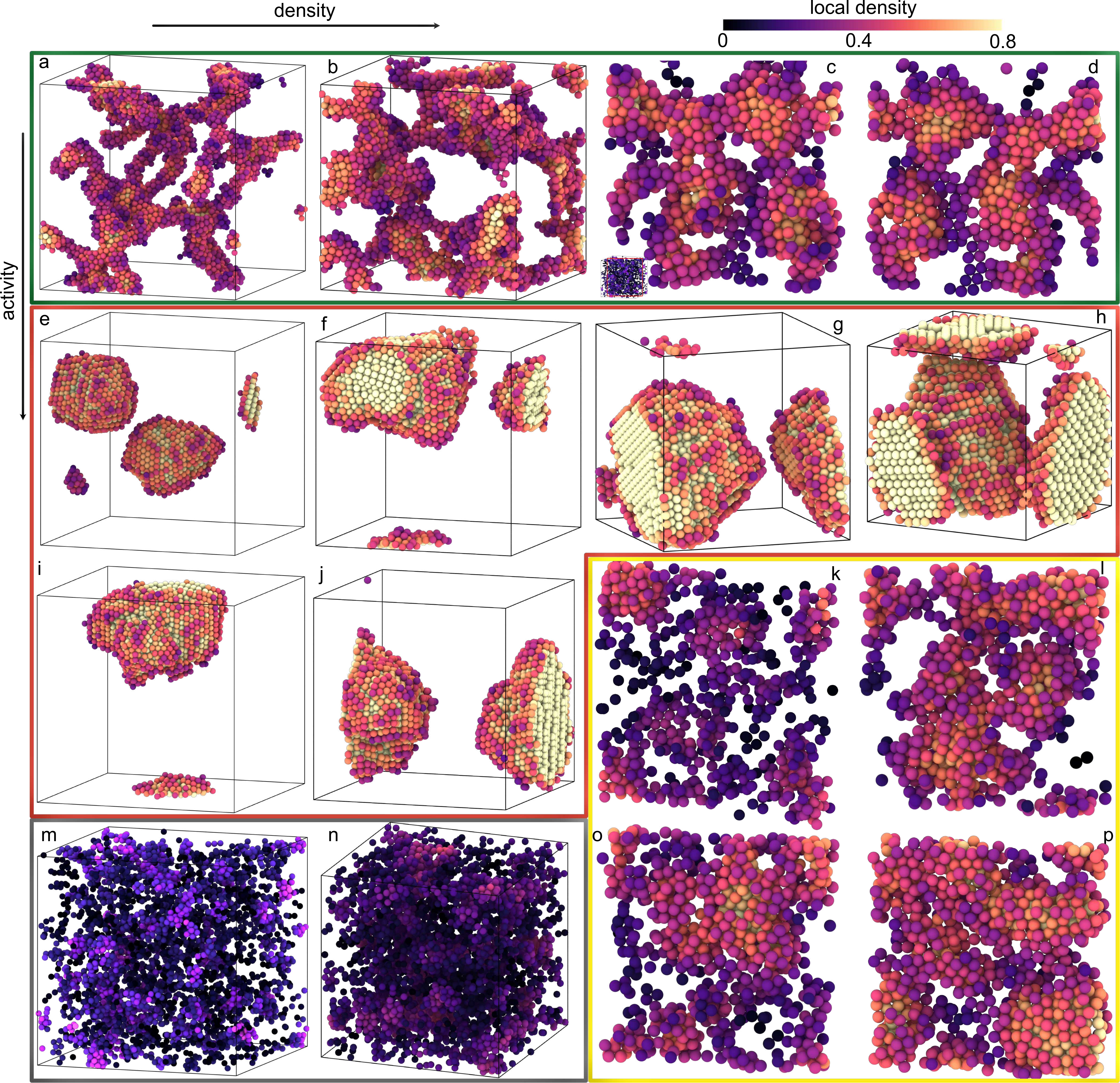}
  \caption{{\bf Phase behavior of attractive active dumbbells.} Snapshots of the dumbbells' system for different global densities $\phi$ and P\'eclet numbers Pe. The attraction $\epsilon=0.3$ for all cases presented here. Particles are colored according to the local density. From top to bottom, the rows correspond to Pe $= 0, \, 50,\, 100, \,150$. From left to right, columns correspond to $\phi=0.05, \, 0.1, \, 0.2, \, 0.3$. At Pe = $0$, panels (a)-(d), the system is in the gel phase. Increasing  activity for the same densities, panels (e)-(h) for ${\rm Pe} = 50$, or for even higher ${\rm Pe} =100$ and small densities (i)-(j), aggregation is favoured and there is complete phase separation, in the sense described in the main text. At Pe = 150 and low density, panels (m)-(n), the structure breaks down into small pieces. At higher densities, panels (o)-(p), the percolating network subsists.
  The bar on top of the plot shows the scale for the local density with which the beads are colored, the scale we use to distinguish the phases, see Fig.~\ref{fig:3} and its discussion, together with the dynamic properties to be presented in Sec.~\ref{dyn}.  
  }
  \label{fig:2}
\end{figure*}

We start by presenting the phase diagram obtained by scanning various values of the packing fraction $\phi$ and the P\'eclet number $\rm{Pe}$. In Fig.~\ref{fig:2}, snapshots are presented for different combinations of these two parameters.
We find that the system exhibits a variety of diverse phases. We start by discussing them at a pictorial level and we later give a quantitative characterization by studying the distribution of local densities, the distribution of cluster sizes, and the dynamics.


\subsection{The structures}

We commence our analysis by setting ${\rm Pe} = 0$ and examining the structures formed at long times for the global densities $\phi=0.05, \, 0.1, \, 0.2, \, 0.3$, as shown in the first row of Fig.~\ref{fig:2}. In the passive limit the dumbbells aggregate in a gel phase, in which particles form a rather static percolating network. This behavior, called gelation~\cite{Poon02,Coniglio04,Zaccarelli2007,Sciortino2008a,Sciortino2008b}, is due to a rapid cooling of an initially disordered conformation which in conjunction with the dumbbells' anisotropy and the short-range attraction induces the formation of such metastable state. The stable state would be instead phase-separated -- a configuration which, nonetheless, cannot be reached since the thermal energy is insufficient for the dumbbells to rearrange into a single compact cluster. 
 {As a matter of fact, rigid dumbbells interacting with a square-well potential undergo a gas-liquid phase separation~\cite{Munao13,Munao14}}


In the second row of Fig.~\ref{fig:2} we consider a higher activity ${\rm Pe}=50$, while maintaining the same densities as in the first row. The dumbbells now aggregate to form a single cluster at any density. We will refer to this state as phase-separated. The resulting cluster exhibits both translational and rotational persistent motion, which we will characterize in detail in Sec.~\ref{sec:single-cluster}.  {Remarkably, an increase in ${\rm Pe}$  induces the dumbbells to reorganize, destroying the gel network and forming a single cluster, thanks to the activity that acts as an additional source of noise.}


When the activity is further increased to ${\rm Pe} = 100$ (third row), an interesting effect is observed upon varying the density. Specifically, at densities $\phi=0.05$ (i) and $\phi=0.1$ (j), we still find phase separation. However, at higher densities $\phi=0.2$ (k) and $\phi=0.3$ (l), the system becomes more disordered, with small clusters that do not grow in size over time. These clusters connect to each other, forming a phase that is similar to the gel observed at vanishing activity but is more dynamic in nature, with the network connections constantly forming and breaking throughout the simulation (see SM  Movie 1)~\cite{movies}. This phase will be referred to as percolating phase, following the nomenclature of Ref.~\cite{prymidis2015} where a similar behavior was found for a system of self-propelled Lennard-Jones  {spheres}.



Finally, at ${\rm Pe} =150$ (fourth row), we notice that at low densities the activity causes the system to become fully disordered (Fig.~\ref{fig:2}(m,n)). In this case, the activity is strong enough to overcome the aggregation force, leading to the cluster's breakup. Conversely, at higher densities, (o) and (p),  we observe again the formation of a percolating network.


Despite the presence of attraction, we have observed four phases occurring in distinct ways. Such findings suggest the existence of a complex phase diagram, which we will now proceed to characterize. 

\subsection{The local densities}

First of all, we reckon that there are no isolated dumbbells once the steady state is reached.
In order to identify the different phases, we developed a method involving the construction of a Voronoi tessellation~\cite{Okabe2000}. Such a tessellation partitions the simulation box into a set of regions --the Voronoi cells-- surrounding each bead such that all points within each cell are closer to the pertaining bead than any other beads. By computing the local packing fraction $\phi_i$ within each region, defined as the ratio between a single bead's volume and its Voronoi cell volume, we obtain valuable information about the local packing and structure of the system. In fact, this quantity expresses how much free space the bead has and thus whether it is free, or located in a cluster.

\begin{figure}[t!]
  \includegraphics[width=\columnwidth]{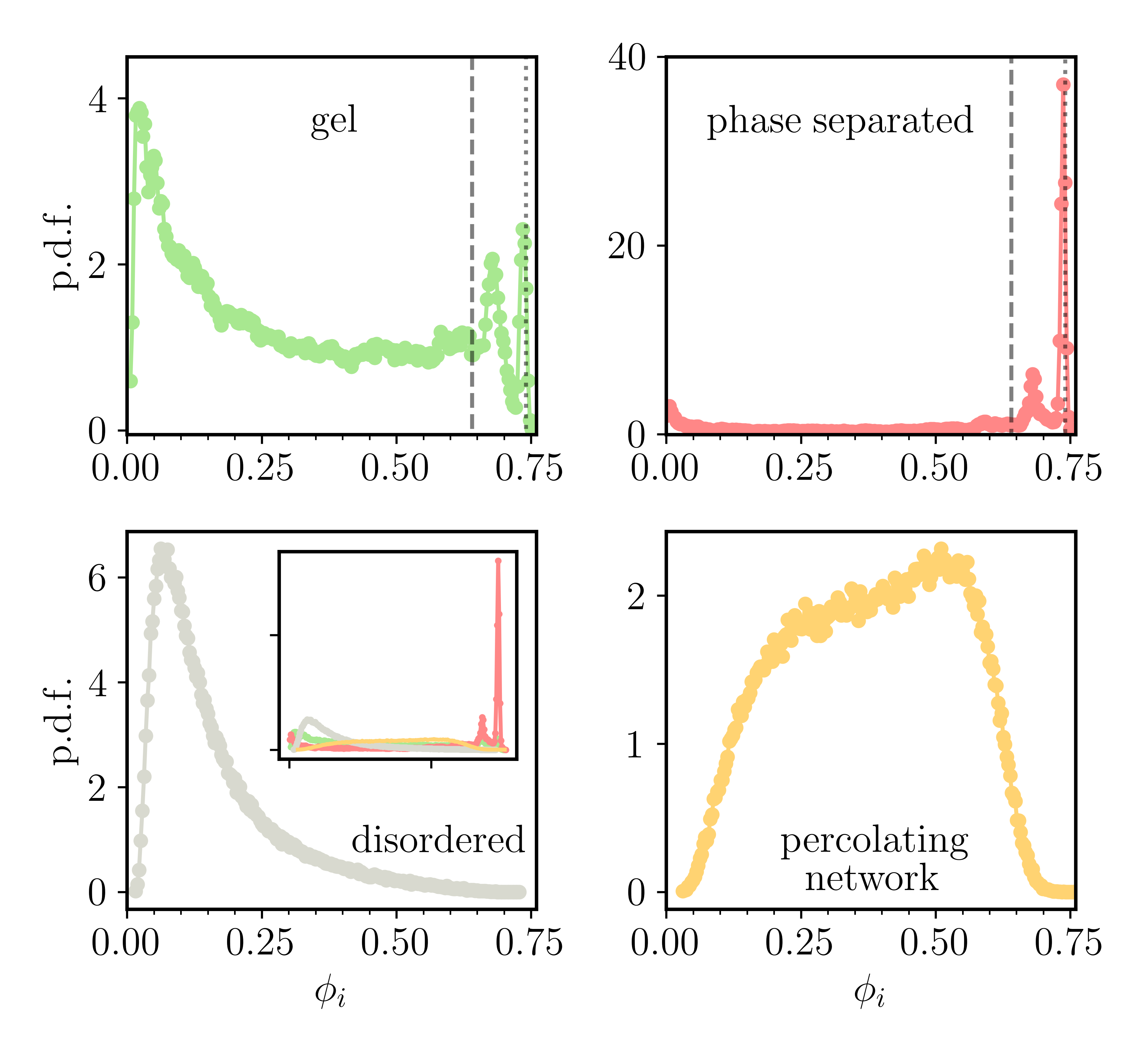}
  \caption{{\bf Local Voronoi density.} Distribution of the local packing fraction of Voronoi cells for different global density and P\'eclet numbers in the four phases:  $\phi = 0.10$ and $\text{Pe}=0$ in the gel,  $\phi = 0.10$ and $\text{Pe}= {50}$ in the phase-separated state,  $\phi = 0.10$ and $\text{Pe}=200$ in the disordered phase, and  $\phi = 0.30$ and $\text{Pe}=150$ in the percolating network. 
  The high density peak in the gel and phase-separated phase 
  are at the close packing value, shown with a dotted vertical line, while the next one close to it is in between $\phi_{\rm rcp} \sim 0.64$ and
  $\phi_{\rm cp}\sim 0.74$, shown with a dashed vertical line, respectively.  {Inset: same distributions plotted together to empathize the relative difference in the peak intensity. The dependence on Pe at fixed $\phi$ and $\epsilon$ is analyzed in Appendix~\ref{app:local-packing}.}
  }
  \label{fig:3}
\end{figure}

In Fig.~\ref{fig:3}, we present the probability distributions of the local density $\phi_i$ in the four distinct phases identified in the study (the global density and ${\rm Pe}$ values are specified in the caption). In each phase the distribution has unique characteristics which can be used as a fingerprint of each phase to classify different configurations. 

In the gel phase, the distribution has three well-defined peaks, one at low and two at high density. The low density peak is due to the beads placed on the border of the gel;
these beads are located inside very large Voronoi cells and hence return a very low $\phi_i$. Instead, the high density peaks are due to beads inside the dense regions of the gel. One is centered around the close packing value $\phi_{\rm cp} \sim 0.74$, while the other is at a value in between the random close packing $\phi_{\rm rcp} \sim 0.64$ 
and $\phi_{\rm cp} \sim 0.74$. We note that for these parameters the low density peak carries more weight than each of the two high density ones. This feature also holds for other values of the parameters corresponding to a gel network.

\begin{figure*}
    \centering
    \includegraphics[width=2.0\columnwidth]{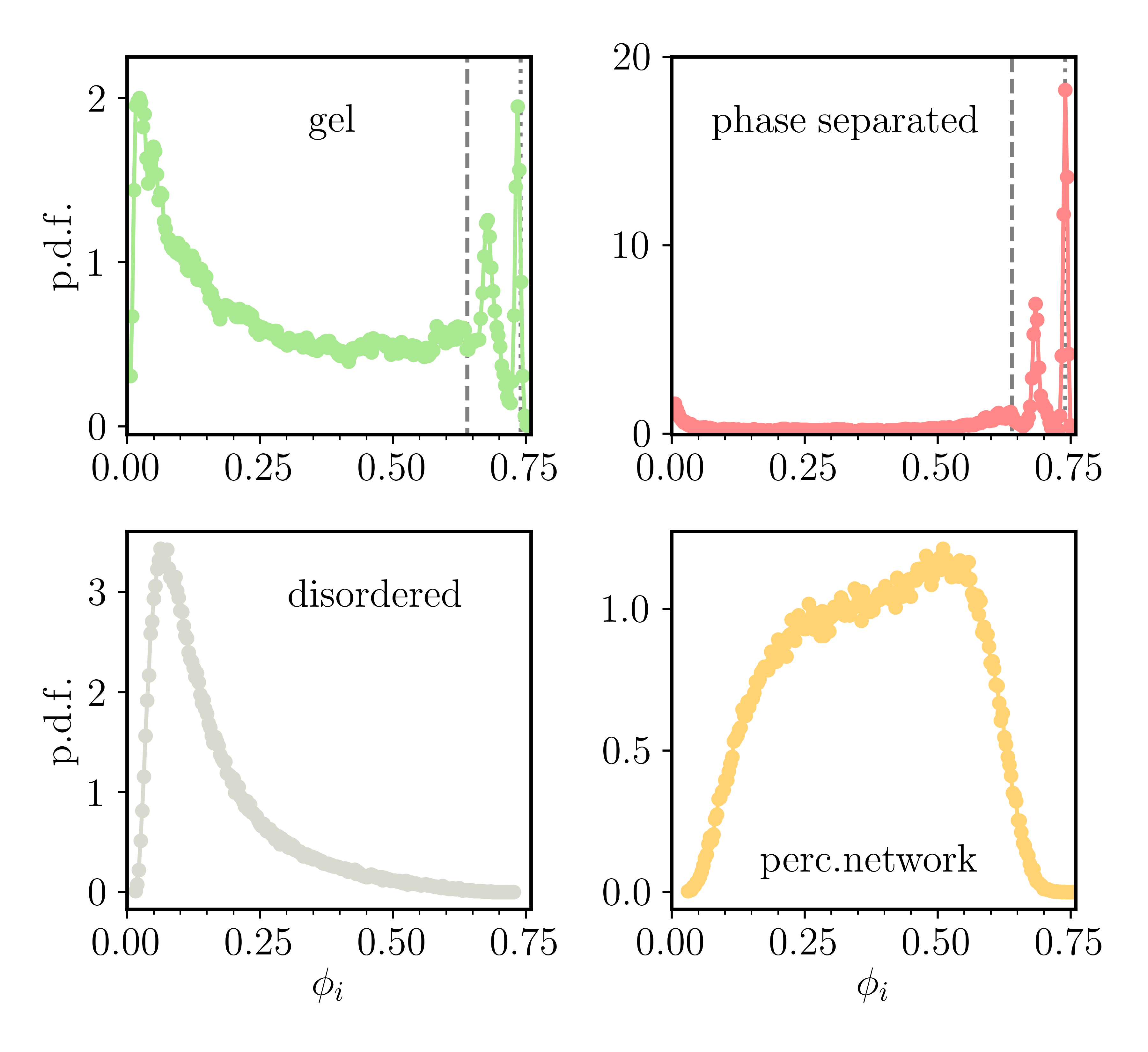}
    \caption{ {\bf Structural analysis of active clusters.} (a)  Snapshot of the system for Pe $=50$ and $\phi = 0.10$ (phase-separated phase), with particles colored accordingly to the number of neighbouring particles inside a cutoff distance of $1.5\sigma$. (b) Snapshot of the same configurations with particles colored accordingly to the common neighbor analysis~\cite{FAKEN1994279}. Particles with FCC order are in blue, those with HCP order in red, while particles with no crystalline order are in grey.
    (c) Distribution of the local packing fraction of Voronoi cells for a configuration in the phase-separated state. The dotted colored lines represent the contribution to the distribution of the beads having a selected Coordination Number (CN), reported in the legend.}
    \label{fig:coord_numb_hist}
\end{figure*}

In the phase-separated systems, the low density peak also corresponds to  {spheres} which are on 
the single cluster surface. At high local densities, there is a double peak structure with one of the peaks sitting on the close packing density and the other one at a slightly lower value, very similar to what we found in the gel. There is, though, an important difference with the gel data, which resides
on the height of these peaks: in the phase-separated phase a much larger portion of particles
have the large densities of these two peaks.

To elucidate the origin of the two peaks in the local density distributions of the gel and phase-separated phase, we analyze the relative frequency of beads with different coordination numbers.
The coordination number of a bead is defined as the number of its nearest neighbours in the Voronoi sense, that is, two beads are considered to be nearest neighbours if they share a common face of the tesselation, and their centers are located within a distance of $1.5 \, \sigma$, Fig.~\ref{fig:coord_numb_hist}(a).
The data in Fig.~\ref{fig:coord_numb_hist} demonstrate that
beads with coordination number 12 contribute to the 
peak at the close packing value. 
More precisely, these beads belong to either a face-centered cubic (fcc) or a hexagonal close-packed (hcp) local crystalline structure in the interior of the cluster.
As shown in Fig.~\ref{fig:coord_numb_hist} (b), the two structures coexist and both have local density $\phi \simeq \phi_{\rm cp}$, consistently with the location of the purple peak in Fig.~\ref{fig:coord_numb_hist} (c).
The secondary peak is associated with beads that have a coordination number equal to 11, and are topological defects that locally disrupt the regular lattice and lower the local packing fraction.
The crystalline structure in the bulk of the cluster resembles very closely the high-density ground state of a system of passive dumbbells, which is a so-called \textit{aperiodic crystal}, with all beads arranged on a close-packed ordered structure and disordered bonds between beads~\cite{Dijkstra_3d-dumbbells}. 
The only difference here is that
we see a mix of fcc and hcp structures. This arrangement is very common in close-packed spheres, and the interfacial planes are called stacking faults. 
We note that our configurations are not  ground state and the system is active, so it is reasonable to expect other close-packed structures to be stable. 
Particles with a lower coordination number contribute to the rest of the distribution and, in particular, to the lower density peak which, as mentioned above, is due to the beads on the surface of the cluster, thus with a smaller number of neighbors compared to the core ones, see Fig.~\ref{fig:coord_numb_hist}(a).

In order to distinguish between the gel and phase-separated phases, it is necessary to check the system's motility: in the gel, dumbbells are almost frozen and cannot diffuse, conversely in the phase-separated case diffusion is observed, see Fig.~\ref{fig:6}. More detailed characterizations of the motility of the full system, and of the single clusters of the phase-separated phase, are given in
Sec.~\ref{dyn} and Sec.~\ref{sec:single-cluster}, respectively.

In the disordered phase, the beads are more homogeneously distributed, resulting in an equally partitioned space and the development of only a single peak at small density, around $\phi_i\sim 0.1$. The percolating network has a significantly different  $\phi_i$ distribution. It is broad, with the appearance of all intermediate densities, with almost equal probability. As noted in the fourth row of Fig.~\ref{fig:2}, the system smoothly transitions between the disordered and the percolating phases.  

 {A comparison between the local packing fraction distributions in the four phases can be appreciated in the inset of the lower left panel in Fig.~\ref{fig:3}. There is a clear difference in the height of the peaks for the four phases.
We also show in Fig.~\ref{fig:dist-pe} in the Appendix~\ref{app:local-packing} how the intensity of the peaks changes with varying Pe, at fixed $\phi=0.1$ and $\epsilon=0.3$. The variation of Pe takes the system 
from the gel, across the phase separated and finally to the percolating network phase. We see no signature of discontinuity in the height of the two peaks at high local packing fraction when crossing the transition lines.
}

\subsection{The cluster size distribution}

\begin{figure}[t]
    \centering
    \includegraphics[width=0.79\columnwidth]{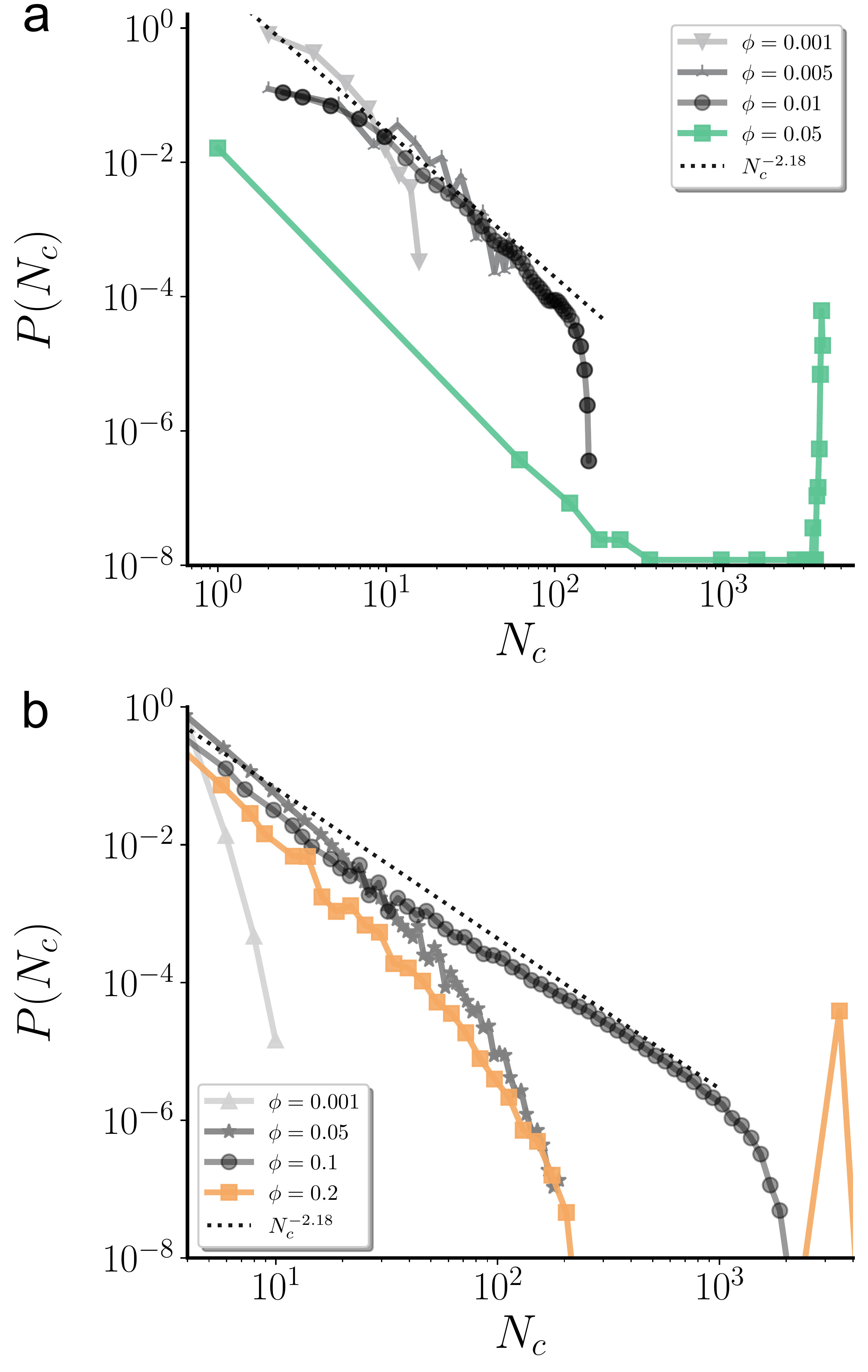}
        \caption{ {\bf Cluster size distribution in the gel, disordered and percolating network cases.} (a) Probability distribution of  cluster sizes $N_c$ at Pe $=5$, for  {four} values of the global packing fraction  {$\phi=0.001, 0.005, 0.01$ (grey curves corresponding to the disordered phase)} and $\phi=0.05$ (green curve corresponding to the gel phase). Panel (b) shows the cluster size distribution for $Pe=150$, and  {$\phi=0.001, 0.05, 0.1$ (grey curves corresponding to the disordered phase)} and $\phi=0.2$ (yellow curve for the percolating network phase).   The dotted line corresponds to an algebraic decay $N_c^{-\tau}$ with Fisher exponent $\tau=2.18$,  { added for comparison}.
        The cluster size distribution was obtained by counting the number of coarse-grained clusters identified with the DBSCAN algorithm~\cite{dbscan, digregorio2019, caporusso2020, negro2022hydrodynamic}. For panel (a) 10 different simulations where performed, and for each of them $10^{3}$ configurations where used for sampling. Panel (b) has been obtained using a single run and $5\times 10^{3}$ configurations, exploiting the shuffling induced by activity.}
    \label{fig:percolation}
\end{figure}

 {In order to identify the global density $\phi$ at which the system percolates, one can proceed as in the usual analysis of percolation~\cite{Stauffer} and gel~\cite{Sciortino2008a,Zaccarelli2007,Sciortino2008b} transitions. A common observable to detect a percolating structure is the cluster size distribution, displayed in Fig.~\ref{fig:percolation} for ${\rm Pe} = 5$ and $150$ and various packing fractions. At both Pe values we see a transition between a disordered phase and a percolating one.
After clustering each conformation using DBSCAN (see the method description in Sec.~\ref{sec:model}), we compute the distribution of cluster sizes. While for sufficiently low densities the distributions have an exponential decay,
near the density where clusters start to connect to form a percolating network, the cluster size distribution takes a power-law form with an exponential cutoff~\cite{Stauffer} ($\phi = 0.01$ for Pe=5 and $\phi = 0.1$ for Pe=150). Notably, the distributions at both the percolating network and gel transitions are compatible with the algebraic law $N_c^{-\tau}$, with $\tau$ the Fisher exponent of standard random percolation for a three-dimensional system, $\tau\sim 2.18$, associated to a fractal dimension 
$d_f\sim 2.53$ by a hyperscaling relation {, see dashed lines in Fig~\ref{fig:percolation}(a,b)}. Increasing density, percolation starts to occur and the system becomes interconnected, with a large cluster and a few small ones. In this case, the cluster size distribution appears to have a peak at a cluster size of the order of the number of beads in the system ($\phi = 0.05$ for Pe=5 and $\phi = 0.2$ for Pe=150). Note that the different ranges of cluster sizes observed in Fig~\ref{fig:percolation}(a,b) stem directly from the different densities considered in the two cases, given that the number of dumbbells remains fixed in both cases.}

\subsection{The phase diagram}

\begin{figure}[t]
  \includegraphics[width=0.9\columnwidth]{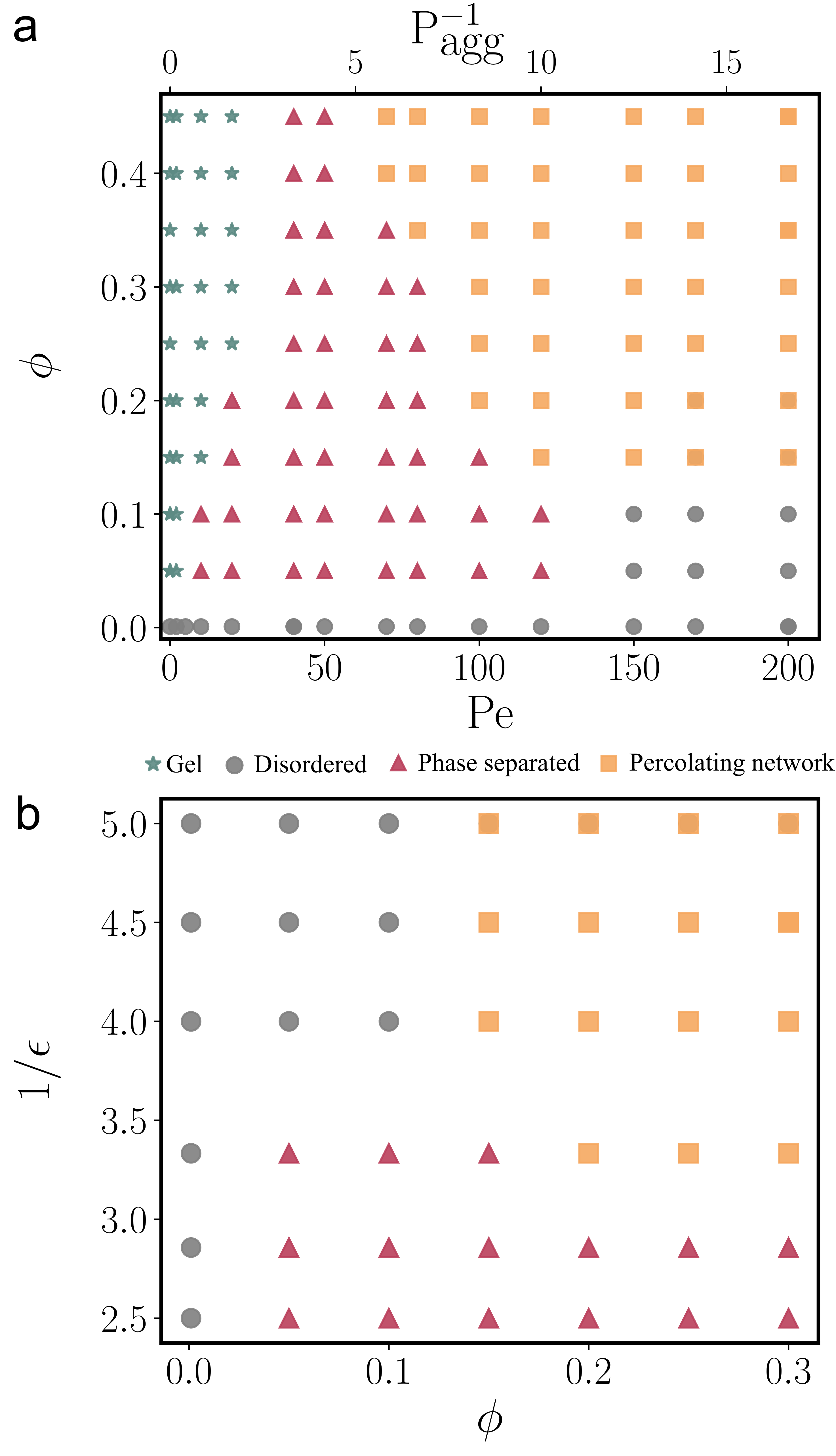}
  \caption{{\bf Bulk phase behaviour.} (a) Phase diagram in the Pe-$\phi$ plane, at fixed $\epsilon=0.3$. (b) Phase diagram in the  {$\phi-1/\epsilon$} plane for Pe $=100$. }
  \label{fig:4}
\end{figure}

The aforementioned observations allowed us to identify at each Pe and $\phi$ the corresponding phase, and build a complete phase diagram, which is displayed in Fig.~\ref{fig:4}a for the particular case $\epsilon = 0.3$.  The analysis of the dynamics which will be presented in Sec.~\ref{dyn} will further support this classification.

Let us fix $\phi$ and discuss the various transitions found 
upon increasing Pe in  {Fig.~\ref{fig:4}a}. 
At $\phi=10^{-3}$ (very dilute system), and for all ${\rm Pe}$, the system appears to be always disordered. Upon considering $\phi=0.05$, we observe a small region where the system is in a gel phase, followed by the appearance at ${\rm Pe} = 10$, or 
$P^{-1}_{\rm agg} = 0.83$, of phase separation  {(the transition is located with our resolution between Pe=2 and 10)}. This is due to the activity acting as an additional noise, which enables particles to rearrange and to not get stuck in a metastable conformation. At ${\rm Pe} = 150$, or $P^{-1}_{\rm agg} = 12.5$, activity is high enough so that the phase-separated configurations start to break, and 
the disordered phase takes over  {(the transition is between Pe=120 and 150)}.
At  a higher packing fraction, {\it e.g.} $\phi=0.15$, we also observe a transition between gel and phase-separation, now at a slightly higher ${\rm Pe}$ or $P^{-1}_{\rm agg}$. We attribute this increase with respect to what was found for lower $\phi$ to the fact that at higher densities the gel becomes thicker, and thus more difficult to break {, see Fig.~\ref{fig:coord-dist} in the Appendix}. At ${\rm Pe} = 120$, or $P^{-1}_{\rm agg} = 10$, we observe a transition between a phase-separated phase and a percolating network  {(the transition is between Pe=100 and 120)}. Again, activity is strong enough to break the single dense phase, but the density is high enough so that the small clusters are interconnected, 
forming the network. At still higher $\phi$,  {we observe the same transitions between phases as the ones observed at $\phi=0.15$. Notably, for increasing global packing fraction, the gel and the percolating network phases tend to squeeze the 
phase-separated regime, reducing the range in Pe where the latter phase occurs. We did not probe densities near close-packing, as it is out of the scope for this article.}

 {We can now explore the effects of the beads attraction strength $\epsilon$. 
These can be visualized either as a new phase 
diagram using as axes $(\phi, 1/\epsilon)$ plane, built at fixed $k_BT$
and Pe, or by changing the value of $\epsilon$ and constructing the same phase diagram as in Fig.~\ref{fig:4}a. The former plot is presented in Fig.~\ref{fig:4}b for Pe=100, while the latter is shown in Fig.~\ref{fig:appendix1} for $\epsilon=0.2,\,\ 0.4$ ($1/\epsilon=5,\,\ 2.5$ respectively). }

 {Looking at Fig.~\ref{fig:appendix1}, we find all 
the four phases described in Fig.~\ref{fig:3}, and a subdivision similar to Fig.~\ref{fig:4}a of the regions pertaining to each phase. However, the critical Pe 
between these phases are shifted. 
In more detail, for $\epsilon=0.2$ the transition between gel and phase-separated at low $\phi$ and the one between phase-separated and percolating network at sufficiently high densities take place at smaller Pe than for $\epsilon = 0.3$.
This can be attributed to the fact that both activity and temperature have a higher role in degrading the clustered system as the ratio Pe$/\epsilon$ is increased. 
Conversely, for $\epsilon=0.4$,  it takes a stronger activity to break the clusters. Both gel and phase-separated phases extend towards slightly higher values of Pe at the same packing fraction.}

 {When slicing the plots of Fig.~\ref{fig:4}a and Fig.~\ref{fig:appendix1} at Pe=100, one can form the plot of Fig.~\ref{fig:4}b. 
Increasing $\phi$ from $\phi=0.01$ up to $\phi=0.3$ we see a disordered phase turning into a phase-separated regime for high $\epsilon$ (or low $1/\epsilon$), where the contribution of attraction is stronger then activity, and into a percolating network for low $\epsilon$ (or high $1/\epsilon$), where instead activity is more important and breaks clusters. Notably, the topology of this phase diagram is very similar to the one in Fig.~2 of Ref.~\cite{Zaccarelli2007} derived for a 
passive system.}

 {Note that the effect of breaking phase-separated configurations through the action of activity was discussed preliminary in Ref.~\cite{schwarz2012}, where it was measured, starting from a passive gel, the threshold value of attractive strength $\epsilon$ needed to break the gel. The authors observed that with activity this threshold increases, meaning that activity enhances breaking of clusters of dumbbells. }



\subsection{Dumbbells dynamics}
\label{dyn}

We now complement the analysis of the structural properties that led us to the phase diagram in Fig.~\ref{fig:4} with a characterization of the dynamics. In particular, we measure the global dumbbells' mean square displacement (MSD) in each of the four phases, and the motion of 
the single clusters formed via phase separation when a non-vanishing active force 
is applied.

\begin{figure}[t]
  \vspace*{0.2cm}
  \includegraphics[width=1.0\columnwidth]{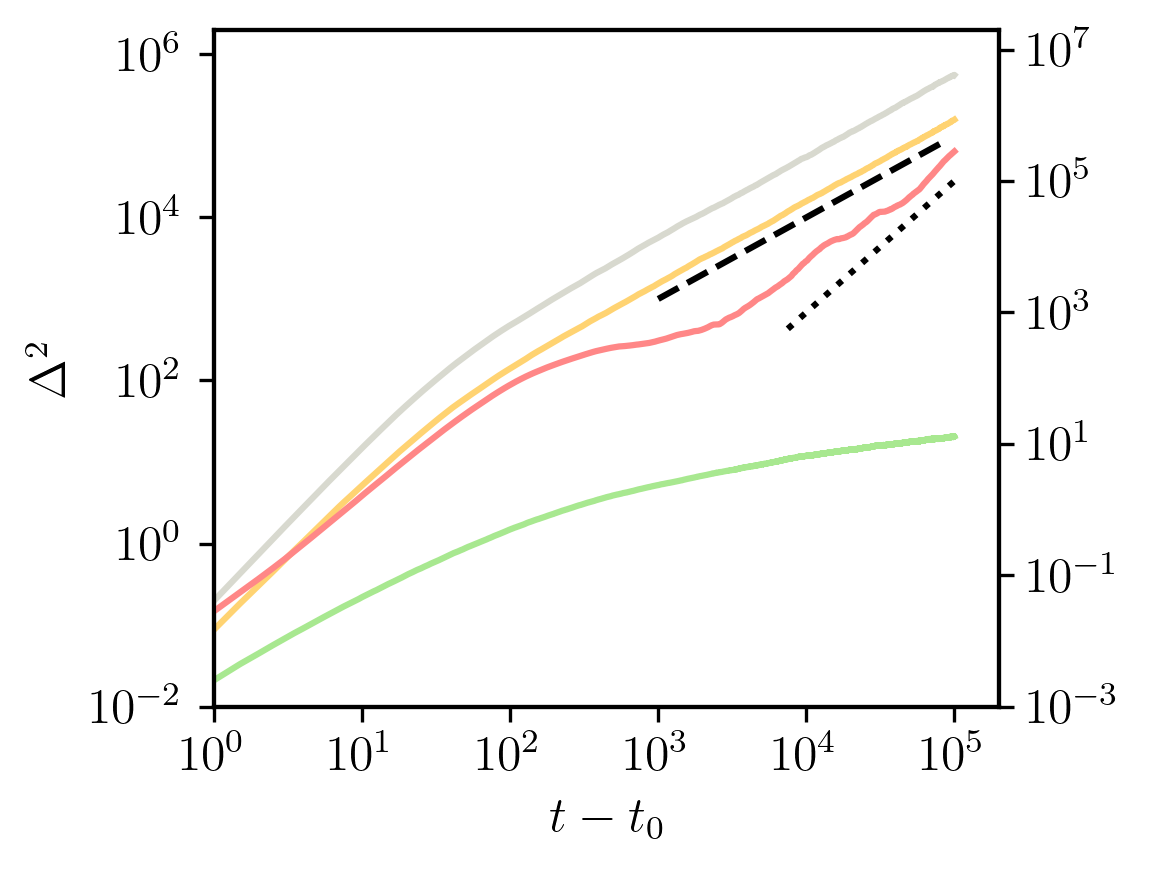}
 \caption{{\bf Mean-square displacement} (MSD) in the different phases.
  Left scale: MSD in the gel (green, Pe = 0 and $\phi = 0.1$), disordered phase (gray, Pe $= 200$ and $\phi = 0.1$) and percolating network (yellow, Pe $= 150$ and $\phi = 0.3$). The percolating network and disordered phases exhibit diffusive behaviour at late time delays, as evidenced by the linear fit (dashed line), while the dynamics of the gel is much slower and may also tend to freeze. Right scale: MSD in the phase-separated case  (red curve)
  with Pe = 50 and $\phi = 0.1$. The dotted line represents a quadratic law, suggesting that the motion of the dumbbell clusters is, on average, ballistic in this case. 
  }
  \label{fig:6}
\end{figure}

The total mean square displacement (MSD) is the result of the sum 
of the MSDs of the $2N$ colloids composing the $N$ dumbbells. It is defined as
\begin{equation}
 {
    \Delta^2(t,t_0)=\frac{1}{2N}\sum_{l=1}^{2N} \, 
    \langle
    ({\bm r}_l(t)-{\bm r}_l(t_0))^2
    \rangle
    \; ,}
\end{equation}
 {where ${\bm r}_l$ are the positions of the centers of the spheres.} 
Fig.~\ref{fig:6} shows $\Delta^2$ for four combinations of parameters $\phi$ and $\rm{Pe}$ representative of the four phases we described beforehand.

In the gel phase, the MSD   is sub-diffusive
(green curve in Fig.~\ref{fig:6}). This can be understood in terms of particles being confined within the gel, resulting in limited movement. In contrast, the disordered phase and percolating network (grey and yellow curves in Fig.~\ref{fig:6}) show diffusive motion after a brief ballistic regime.


When considering the phase-separated system (red curve in Fig.~\ref{fig:6}), 
there is, instead, a ballistic behaviour at long time-scales, characterised by a quadratic behaviour $\Delta^2 \sim t^2$, associated to the motion of a single aggregate, as we do not see any particles in the dilute phase. This indicates that the cluster exhibits persistent motion over time, and suggests to analyze its dynamics in more detail.

 {Note that in these plots we do not explicitly subtract the global motion of the center of mass of the system. The results are unaffected by it, except for the phase-separated system, where we observe ballistic motion of a single cluster. Here, subtracting the motion of the center of mass, which coincides with the cluster's center of mass, means not accounting for its overall characteristic motion.}

\section{Single cluster motion}
\label{sec:single-cluster}

We now characterize the ballistic motion of a single cluster in the phase separated regime. We use a  kinematic and dynamical approach to decipher the interplay between frictional dissipation 
and active forcing which leads to the persistent motion of the active  cluster.

\begin{figure}[t!]
    \centering
    \includegraphics[width=\linewidth]{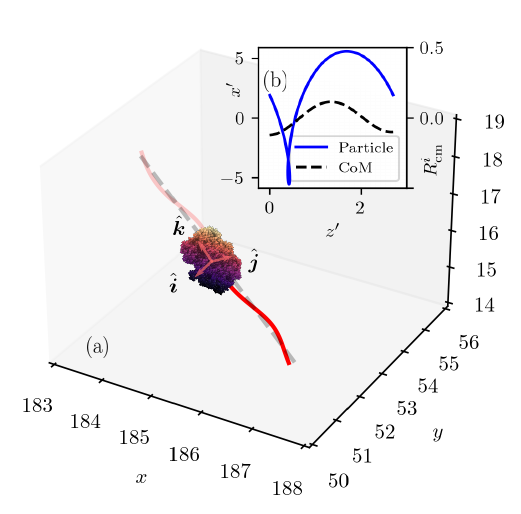}
    \caption{{\bf Approximately helical trajectory of a typical cluster.} (a) In red the 
    trajectory of the center of mass of a cluster with $N_c=3728$ beads, with no surrounding gas, formed by aggregation at $f_{\rm{act}} = 1$ and evolving at $T=0$. The size of the 
    cluster has been reduced by a factor of $2.5$ to ease the visualization of the trajectory. 
    The scales of the Cartesian axes are measured in units of the bead's diameter $\sigma$.
    The aggregate is, approximately, an ellipsoid of revolution, or spheroid. The approximate lengths of the semi-axes of the associated inertia tensor are $(10.11, \, 9.71, \, 13.34)$. The principal eigenvector is roughly parallel to $\hat{\bm{k}}$ -- the direction along which the net force momentum is null -- while the remaining eigenvalues associated to the other 
    two eigenvectors are roughly degenerate, taking values $(17.7, \, 16.2, \, 37.5)$. 
    In dashed gray the axis of the helix.
    The $\hat{\bm{i}}, \, \hat{\bm{j}}, \, \hat{\bm{k}}$ Cartesian and inertial 
    coordinate system, superimposed to the cluster, moves along the helical axis with constant velocity. The unit vectors do not change their orientation with respect to the laboratory reference frame, 
    with $\hat{\bm{k}}$ oriented in the direction of the helical 
    axis and $\hat{{\bm i}}, \, \hat{{\bm j}}$ and perpendicular to it.
    (b) One period of the trajectories of the center of mass (dashed line, right scale), 
    and of one bead in the cluster (blue line, left scale), on the 
    $\hat{\bm{k}}, \, \hat{\bm{i}}$ plane.
    Two movies in the SM highlight these two trajectories in  the full three-dimensional space.  
    }
    \label{fig:cluster_dyn}
\end{figure}

Figure~\ref{fig:cluster_dyn}(a) shows a typical trajectory of the center of mass (CoM)
of a single cluster (red curve). This trajectory was obtained by extracting one typical cluster from the 
system and placing it in a box with no gas dumbbells around (in the figure the cluster has been resized to 
make the trajectory visible).
More details of the motion can be appreciated in the 
video~SM Movie 2. The cluster moves along a trajectory resembling a helix. Indeed, its motion can be seen as the composition 
of a linear motion at constant velocity along the direction defined by the axis of the helix
(dashed grey straight line), and an additional rotation around this axis. 
 {Notice that 
the radius of the helix is of the order of the beads size; therefore, rotational motion around the persistence axis does not significantly contribute to the MSD, as it will be rationalized in the following.}
Importantly, once the cluster is formed, the beads do not change their arrangement, so that the cluster effectively behaves as a rigid body.

We introduce a new reference frame, with its center placed on a fixed point within the 
cluster and moving along the axis of the helix with the same persistence velocity of the cluster itself.  {The helix axis is found as the best fit of the trajectory of the cluster center of mass with a straight line.}
The axes of this new coordinate system are defined by the unit vectors $\hat{\bm{i}}$, $\hat{\bm{j}}$, $\hat{\bm{k}}$ and are chosen in such a way that $\hat{\bm{k}}$ is the direction of the helical axis and $\hat{\bm{i}}$, $\hat{\bm{j}}$ are mutually perpendicular so that the triad forms an orthonormal basis for the three-dimensional space. This is shown in Fig.~\ref{fig:cluster_dyn}(a).
We stress that such a reference frame is inertial, as it only translates (and does not rotate) with respect to the laboratory frame with constant velocity.

In order to rationalize the dynamics of the cluster, we plot in 
Fig.~\ref{fig:cluster_dyn}(b) the CoM trajectory during a single helical period 
on the $ \hat{\bm{k}}, \hat{\bm{i}}$ 
plane  (dashed black line), along with the motion of a generic bead in the cluster (solid blue curve). 
Two movies in the SM give a three-dimensional  {view} of these motions. 
By comparing the two, one finds that their motion is periodic with the same period.
However, while the CoM trajectory has a sinusoidal profile (with a rather small amplitude), 
consistently with a circular motion on the 
$ \hat{\bm{i}}, \hat{\bm{j}}$ plane normal to the helix axis,  the bead's motion describes an epicycloid.
This is the fingerprint of the composite dynamics of the beads which rotate around the CoM while the latter rotates around the helix axis. 
Such dynamics is reminiscent of that of the Moon orbiting the Earth and, in turn, rotating on itself with the same period. Analogously, the cluster rotates around the helix axis showing always the same face (see SM Movie 3). 

We stress that the motion displayed by the cluster in Fig.~\ref{fig:cluster_dyn}  is not a particular case;
it is consistently observed for any typical cluster in the phase-separated region of the phase diagram. 
We have checked that the analysis that we develop below describes the motion of such clusters with number of 
beads ranging from 100 to 4000. Other clusters with more complex forms and dynamics can also exist (as the 
result, for example, of the aggregation of two colliding ones) but we do not discuss them here. 

\begin{figure*}[!ht]
    \centering
    \includegraphics[scale=0.9]{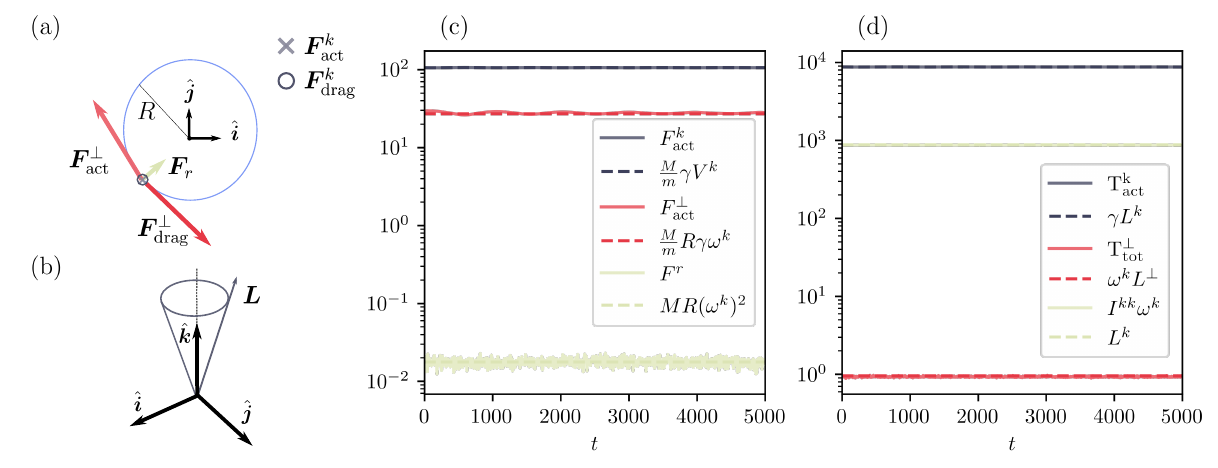}
    \caption{{\bf Schematic representation of the forces on the cluster and its helical and precessional motion.} (a) Top view, showing the components of  ${\bm{F}}_{\mathrm{act}}$ and ${\bm{F}}_{\mathrm{drag}}$ on the  {plane normal to $\hat{\bm{k}}$} , as well as the radial force ${\bm{F}}_r$. (b) Side view, showing the precessional motion of the angular momentum $\bm{L}$ around the axis $\hat{\bm{k}}$.  {Unit vectors $\hat{\bm{i}}$, $\hat{\bm{j}}$ in panels (a) and (b) represent orthogonal basis vectors defining the plane normal to $\hat{\bm{k}}$.} (c)  {Time evolution of  $F^k_{\mathrm{act}}$, the $k$ component of the active force $\bm{F}_{\mathrm{act}}$,  of $F^{\perp}_{\mathrm{act}}$, the modulus of $\bm{F}_{\mathrm{act}}$ perpendicular to $k$, and of $F^r$,} compared to the theoretical predictions (dashed lines) from the equations described in the text. (d) Time evolution of the active torque  {${\bm T}_{\mathrm{act}}$} and angular momentum $\bm{L}$, also compared to the theoretical predictions (dashed lines). 
    }
    \label{fig:11}
\end{figure*}


To go beyond the kinematic description of the cluster's motion, we now solve its dynamics building on the assumption that the cluster is, for all practical effects, a rigid body. Therefore, the motion can be described by separating positional and rotational degrees of freedom.

The CoM motion is governed by the effects of the total active force and friction.
Newton's equation for the CoM position ${\bm{R}}_{\rm cm}$ reads
\begin{equation}
M\ddot{\bm{R}}_{\rm cm} = \bm{F}_{\rm drag} + \bm{F}_{\text{act}}
\; ,
\end{equation}
with $M=mN_c$ the total mass of the cluster, 
$N_c$ the number of particles in the cluster, 
${\bm{R}}_{\rm cm}=N_c^{-1}\sum_{l=1}^{N_c}{\bm{r}}_l$ the position of the CoM, 
$\bm{F}_{\text{act}} = \sum_{l=1}^{N_c} {\bm{f}_{\text{act}}}_l$ the total active force, and 
$\bm{F}_{\rm drag}=-(M/m)\gamma \dot{\bm{R}}_{\rm cm}$ the total drag acting on the center of mass. 
We stress that internal forces, 
arising from the Mie potential and bond constraints, do not contribute to the dynamics of the center
of mass.

The equations describing the rotational dynamics are given by
\begin{equation}
    \dot {\bm L} =  {\bm{T}_{\rm drag} + \bm{T}_{\rm act}}
    \; .
    \label{eqrot}
\end{equation}
Here ${\bm L}= \sum_{l=1}^{N_c} \bm{r}'_l \times \dot{\bm{r}}'_l$ is the angular 
momentum of the cluster computed choosing the CoM as  {reference point}, 
with $\bm{r}'_l=\bm{r}_l-\bm{R}_{\rm cm}$ the position of the $l-$th bead with respect to the CoM. 
The right-hand-side is the total force momentum,  
decomposed in a contribution from the active force  $ {\bm{T}_{\rm act}}=\sum_{l=1}^{N_c} \bm{r}'_l\times {\bm{f}_{\rm act}}_{l}$ and another one from the drag force $ {\bm{T}_{\rm drag}}=-\gamma {\bm L}$.
Once more, torques originating from internal forces do not contribute to the overall rotational dynamics. 

Before proceeding, we stress that a peculiarity of this system is that the active force $\bm{F}_{\text{act}}$ and  {torque $\bm{T}_{\rm act}$} are  {body-fixed}, 
so that they undergo the same rotational and translational motion as the rigid body, as beads do not reposition during the trajectory.

With the equations of motion at hand, we are now ready to derive an expression relating forces and  {torques} with kinematic quantities. 
First, we observe that the CoM acceleration is null along 
the $\hat{\bm{k}}$ direction, therefore the $k$-th component of the active force is counterbalanced by the drag:
\begin{equation}
F_{\rm{act}}^{k}=\dfrac{M}{m}\gamma V^k
\; ,
\label{eqn:FactK}
\end{equation}
where $\bm{V}=\dot{\bm{R}}_{\rm cm}$, so that, by construction, $V^k$ is the persistence velocity of the cluster along the helix axis, measured in the laboratory reference frame.

{  
Conversely, in the  {perpendicular plane defined by the unit vectors} 
 {$ \hat{\bm{i}},\hat{\bm{j}} $}, the CoM velocity 
rotates with constant angular velocity $\bm{\omega}=\omega^k \hat{\bm{k}}$, see Fig.~\ref{fig:11}(a). 
As the dynamics of the cluster is also in the overdamped regime, the active force is approximately counterbalanced by the frictional force so that to $\bm{F}_{\rm{act}}^{ {\perp}} \approx M\gamma \bm{V}^{ {\perp}}/m$, with the mismatch $\bm{F}^{r}=\bm{F}^{\perp}_{\rm drag} + \bm{F}^{\perp}_{\text{act}}$ between the active and the frictional force directed radially with respect to the circle in the normal plane. This radial force becomes a non-negligible inertial contribution, that is never counterbalanced by drag and acts as a centripetal force pulling the cluster towards the axis of the helix. It is indeed the combination of the balancing of the forces along the tangent to the cluster trajectory and the centripetal unbalanced radial component that gives rise to the helical motion. Moreover, since  the circular motion occurs with uniform velocity $V^{ {\perp}}=\omega^k R$, with $R$ the radius of the circle, the forces can be expressed in terms of the rotational velocity, as follows:
}
\begin{align}
F_{\rm{act}}^{ {\perp}}&=\dfrac{M}{m}R\gamma\omega^k \; , 
\label{eqn:FactT} 
\\
F^{r}&=MR{(\omega^k)}^2 \; . \label{eqn:FactR}
\end{align}
(Note that in Eq.~(\ref{eqn:FactT}) we have dropped the inertial contribution, consistently with the observation that $\omega^k$ does not change in time.) 


We now test whether the formul\ae \, (\ref{eqn:FactK})-(\ref{eqn:FactR}) are in agreement with the simulation data. 
In Fig.~\ref{fig:11}(c), we show the forces considered (continuous lines) as functions of time, and we compare them to the corresponding time averaged quantities appearing on the right-hand-sides of Eqs.~(\ref{eqn:FactK})-(\ref{eqn:FactR}) (dashed lines). We find good agreement in all cases. In particular, we notice that the $\hat{\bm{k}}$ direction roughly coincides with that defined by the principal eigenvector of the inertia tensor $\bm{I}$ while, instead, the radial force $F^r$ is significantly smaller than the other ones. Moreover, the forces are hierarchically organized, with $F_{\rm act}^k \sim 100$, $F_{\rm act}^{ {\perp}} \sim 20$, and $F^r \sim 10^{-2}$, 
in units of $f_{\rm act}$.

We now proceed with the analysis of the rotational dynamics by assuming that the angular momentum component $L^k$ in the $\hat{\bm{k}}$ direction remains constant over time -- or equivalently drag and active  {torques} compensate along $\hat{\bm{k}}$, i.e. $\gamma L^k= {T_{\rm act}^k}$.  {The component of the angular momentum normal to $\hat{\bm{k}}$ is} constant in modulus, and rotates with angular velocity $\bm{\omega}$ around the helix axis under the action of the total  {torque} $ {\bm{T}_{\rm tot} = \bm{T}_{\rm drag} + \bm{T}_{\rm act}}$.
Under these assumptions, it is straightforward to rewrite the time derivative in Eq.~\eqref{eqrot} so that
\begin{equation}
\bm{\omega} \times \bm{L} =  {\bm{T}_{\rm tot}}
\; .
\label{eqn:precession_L}
\end{equation}
This expression suggests that $\bm{L}$ performs a precessional motion around the helical axis under the action of  {the torque}, see also Fig.~\ref{fig:11}(a).
Moreover, one can also relate the angular momentum along $\hat{\bm{k}}$ with the $kk$ component of the inertia tensor so that 
\begin{equation}
    L^k=I^{kk}\omega^k
    \; ,
    \label{eqLk}
\end{equation}
being $I^{kk}$ the helical axis component of the inertia tensor $\bm{I}$ of the cluster.


These equations are tested against simulations in Fig.~\ref{fig:11}(d), and we find again good agreement. This confirms that the angular momentum performs a precession around the $\hat{\bm{k}}$
axis with precession angular velocity equal to that of the cluster, confirming \emph{a posteriori} our initial assumption. 

\begin{figure}[t!]
    \centering
    \includegraphics[width=1.0\linewidth]{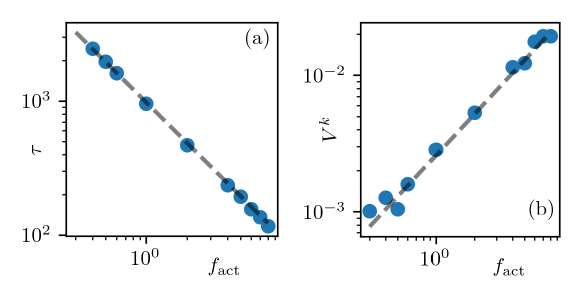}
    \caption{{\bf Motion of a cluster} with mass  $M=3728$ at temperature $T=0$.
    (a) Rotation period $\tau$  
    as a function of the active force $f_{\rm act}$ on a single bead. The slashed straight line represents the theoretical prediction $\tau \propto f_{\rm act}^{-1}$ in Eq.~(\ref{eq:tau}), with $\kappa = 48.71$ and $R=0.204$ measured independently. (b) Linear velocity of the CoM, $V^k$, as a function of the single bead 
    active force $f_{\rm act}$.  The dashed line is the theoretical prediction inverting Eq.~(\ref{eqn:FactK}), $V^k = m\ell/(M\gamma) f_{\rm act}$, using $\ell = F^k_{\rm act}/f_{\rm act} \approx 96.34$ (measured independently) 
    and $\gamma = 10$. 
    }
    \label{fig:enter-label}
\end{figure}

As further tests, we checked the dependence of the period on the magnitude of the active force $f_{\rm act}$ derived from Eq.~(\ref{eqn:FactT}),
\begin{equation}
\tau=\frac{2\pi M\gamma R}{\kappa m} \, \frac{1}{f_{\rm{act}}}
\; , 
\label{eq:tau}
\end{equation}
  where $F_{\rm{act}}^{ {\perp}} = \kappa f_{\rm{act}}$ with $\kappa$ a proportionality coefficient measured from simulation data 
  that turns out to be independent of the active force $f_{\rm act}$. 
To this purpose, we considered the same cluster and we varied $f_{\rm act}$. The results of this analysis show that the period and the active force are indeed inversely proportional, see Fig.~\ref{fig:enter-label}(a),
with the proportionality coefficient $\kappa$ depending on the structural properties of each individual cluster considered, 
and being consistent with the parameter dependencies in Eq.~(\ref{eqn:FactT})
At the same time, we find that the radius $R$ remains constant while we 
vary $f_{\rm act}$ (not shown). In fact, inverting Eq.~(\ref{eqn:FactT}), 
one finds that $R=F_{\rm{act}}^{ {\perp}}m/(M\gamma\omega^k)$, with both $F_{\rm{act}}^{ {\perp}}$ 
and $\omega^k$ proportional to $f_{\rm{act}}$, leaving no dependence on the active force.


Finally, we put Eq.~(\ref{eqn:FactK}) to the numerical test. We show in Fig.~\ref{fig:enter-label}(b)
that $V^k$ is proportional to $f_{\rm act}$ with a proportionality constant which agrees with the 
parameter dependence of the prefactor in Eq.~(\ref{eqn:FactK}).

We also checked that the direction of motion ${\bf \hat k}$ does not change in time.

 {We can now comment more about the overdamping assumption, and the presence of a non-negligible inertial contribution that causes the cluster to feel a centripetal force and experiencing the helical motion. This assumption is in line with considering the cluster's rotational motion with a constant angular velocity. The latter, in fact, arises from the drag torque counterbalancing the active torque. This same movement causes the active force to have a body-fixed rotation with the cluster; thus at each timestep this same force is slightly rotated with respect to the counterbalancing drag. It is this mismatch that causes an inertial contribution to appear. This same contribution shapes in fact the motion of the center of mass of the cluster. For instance, we can notice the dependence of $R$ on the damping coefficient $\gamma$. This formula implies that under weaker damping coefficients the 
oscillatory motion around the propulsion axis has a larger amplitude. This is indeed what we find, see Appendix~\ref{app:trajectory}. In fact, smaller damping means a larger force mismatch $\bm{F}^{r}$ and a larger centripetal force due to acceleration. So larger dampings $\gamma$ make the trajectory to look more and more like a straight line.}


We stress that the description of the cluster's dynamics is based on the assumption that the rotational velocity $\bm{\omega}$ is oriented parallel to the helical axis, with negligible transversal components. The latter could lead in principle to more complex effects -- such as nutation for instance -- which, nevertheless, were not observed in simulations\footnote{ {Note that the precession of $\bm{L}$ is caused by $\bm{L}$ not being on the same direction as $\bm{\omega}$ and $\hat{\bm{k}}$ ( the latter parallel to each other). If $\bm{\omega}$ and $\hat{\bm{k}}$ are not parallel anymore, we could see changes in direction of $\bm{\omega}$ as well, making the motion much more complex.}}. 
This also resonates with the fact that in absence of net torques, a rigid body with two degenerate eigenvalues in its inertia tensor would rotate by keeping constant the component of the angular momentum parallel to the non-degenerate eigenvector~\cite{Goldstein}. Interestingly enough, in our case the particle aggregates indeed exhibit elongated shapes with the principle eigenvector roughly parallel to $\hat{\bm{k}}$  {(the product of the two unit vectors is 0.955)}
and the two dimensions being roughly similar, see the values given in the caption of 
Fig.~\ref{fig:cluster_dyn}.
Nevertheless, the motion of the cluster considered here 
is in general significantly more complicated as, in this case, the net force and torque are not null, leading to a more complex dynamics, with the radial force sustaining the rotational dynamics of the CoM of the cluster, ultimately leading to the intriguing helical motion described above. 

 {An interesting point to discuss is the possible effects in the the cluster's motion due to  temperature, or equivalently to Pe. If we fix the dumbbell's arrangement inside the cluster, and if we are at $\phi$ and Pe values where phase-separation is observed, the only effect of temperature would be to add thermal noise, which changes the persistence direction over time. At the same time, thermal noise might impact the arrangement of dumbbells inside a cluster, starting from a disordered conformation, and this in turn could change the total active forces and torques acting on the cluster. In this particular cluster, we observe that particles more likely point in a direction perpendicular to $\hat{\bm{k}}$ (not shown). However, we expect that increasing N, and thus the size of the cluster, that the arrangements of dumbbells would be on average random on all directions. }

 {Another interesting point is to compare the flocking effect found in \cite{Caprini2023} with our system where we do observe persistence, by measuring the parameter $p_c=\frac{1}{N}\sum_{i=0}^N \frac{\mathbf{\dot r_i}}{|\mathbf{\dot r_i}|}$. Although we see $p_c$ peaking where persistent clusters appear (not shown), the magnitude of this parameter is much less than one. The reason behind this is that a portion of $\mathbf{\dot r_i}$ is dedicated to the rotational motion, obscuring the effect of flocking. Moreover, while attractive ABPs can freely rotate, the direction of the dumbbells inside a cluster is fixed in time, changing the nature of the cluster itself.}

\section{Conclusions}
\label{sec:conclusions}

To conclude, we investigated the nature of the phase behavior and the dynamics of an active dumbbell system with attractive interactions in three dimensions. We characterized the phase diagram
in the (Pe, $\phi \leq 0.45$) plane at fixed $\epsilon = 0.3$ 
and we elucidated the effect of the 
strength of the attraction $\epsilon$ by studying the stationary state reached at
Pe = 100 and parameters in the ($\phi \geq 0.2, \, 1/\epsilon \leq 5$) plane.  
In this way we showed that four cases are realized: 
a disordered state, a percolating network, phase separation and 
an active gel. 
 {We believe that this phase diagram will serve as a reference for further works on elongated self-propelled particles in $3D$, and also more realistic models that may include hydrodynamic interactions, which are known to play a highly non trivial role in competition with particles' anisotropy and self-propulsion~\cite{theersHI,negro2022hydrodynamic}.}

Next, we focused on characterising the motion of a typical dense cluster in the phase-separated phase. First, 
we found that these clusters typically take a spheroid form, and for $f_{\rm act} < 10$  displace with constant velocity in a 
direction which is very close to its main axis of symmetry while performing a rotational motion of
very small radius in the transverse plane. All in all, the motion is very close to 
helical.  With some simple arguments, explained in Sec.~\ref{sec:single-cluster}, we then related the 
linear and angular velocities to the strength of the active force acting on the single 
molecules finding very good agreement with the numerical measurements. 
The dynamics of formation of these clusters, a full characterization of their morphology and statistics, and 
many other details are very interesting but fall beyond the scope of this work.

\section*{Acknowledgements}
This numerical work was carried out on the Dutch national e-infrastructure with the support of SURF through Grant 2021.028 for computational time (L. N. C. and G. N.) and on ReCas HPC-Cluster in Bari (Italy).  We acknowledge funding from MIUR Project No. PRIN 2020/PFCXPE and ANR-20-CE30-0031.
We thank P. Royall and F. Turci for very useful discussions.



\appendix
\label{sec:appendix}

\renewcommand\thefigure{\thesection.\arabic{figure}}    
\setcounter{figure}{0}    

\section{Pe influence on the local structure}
\label{app:local-packing}

\begin{figure}[h!]
    \centering
    \includegraphics[width=0.8\linewidth]{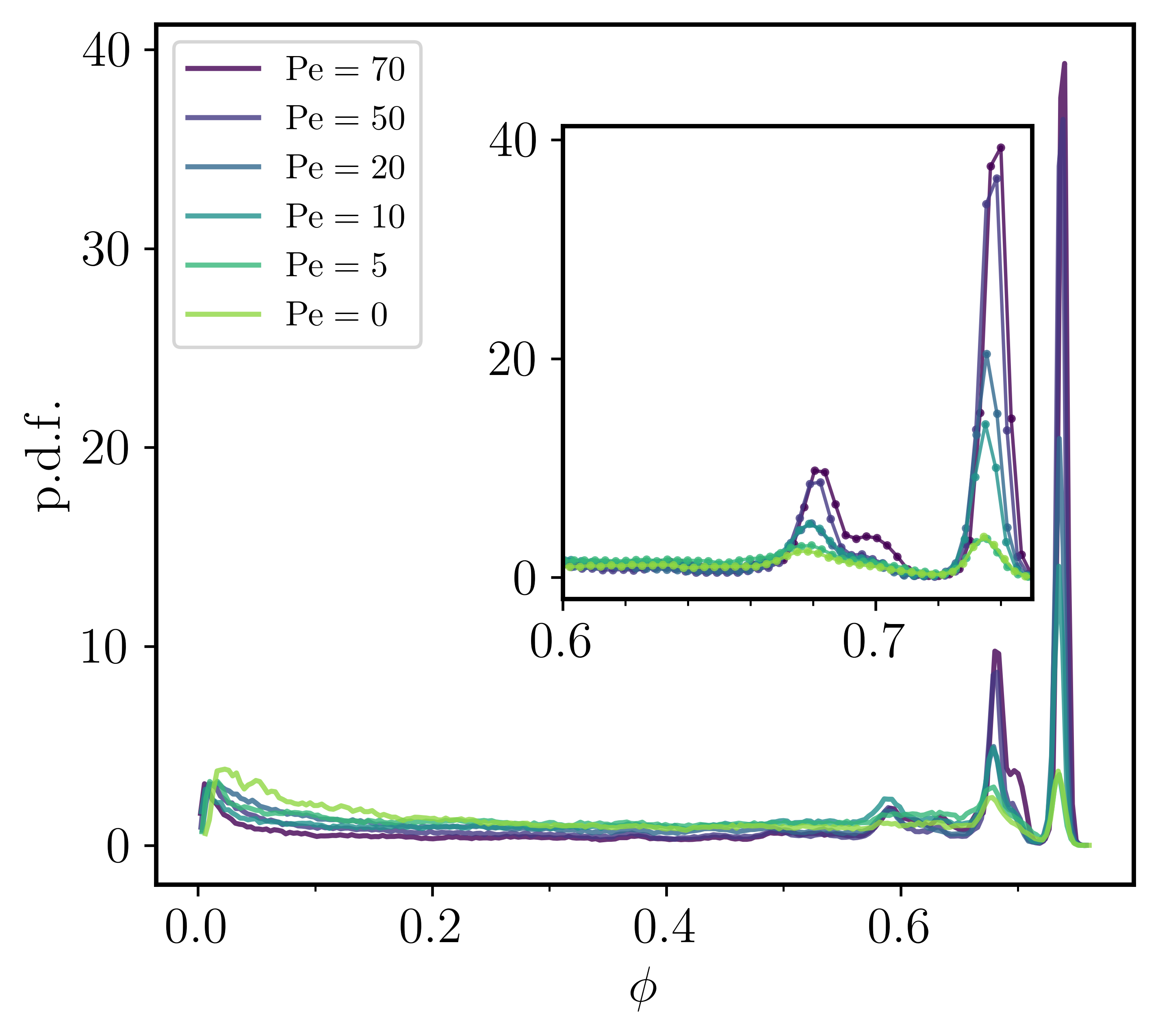}
    \caption{ {Distribution of the local packing fractions of the Voronoi cells at fixed density $\phi = 0.10$ and different P\'eclet numbers reported in the legend. The attraction scale is $\epsilon=0.3$.}}
    \label{fig:dist-pe}
\end{figure}

 {
In Fig.~\ref{fig:dist-pe} we display the 
dependence on Pe of the distribution of local packing fractions of the Voronoi cells  at fixed global 
density $\phi = 0.10$ and $\epsilon = 0.3$. The position of the two peaks at high $\phi$ remains unmodified but their heights change with Pe. More precisely, the two high density peaks get higher and higher as Pe increases. The variation is gradual with no signature of discontinuity.}

\begin{figure}[h!]
    \centering
    \includegraphics[width=0.8\linewidth]{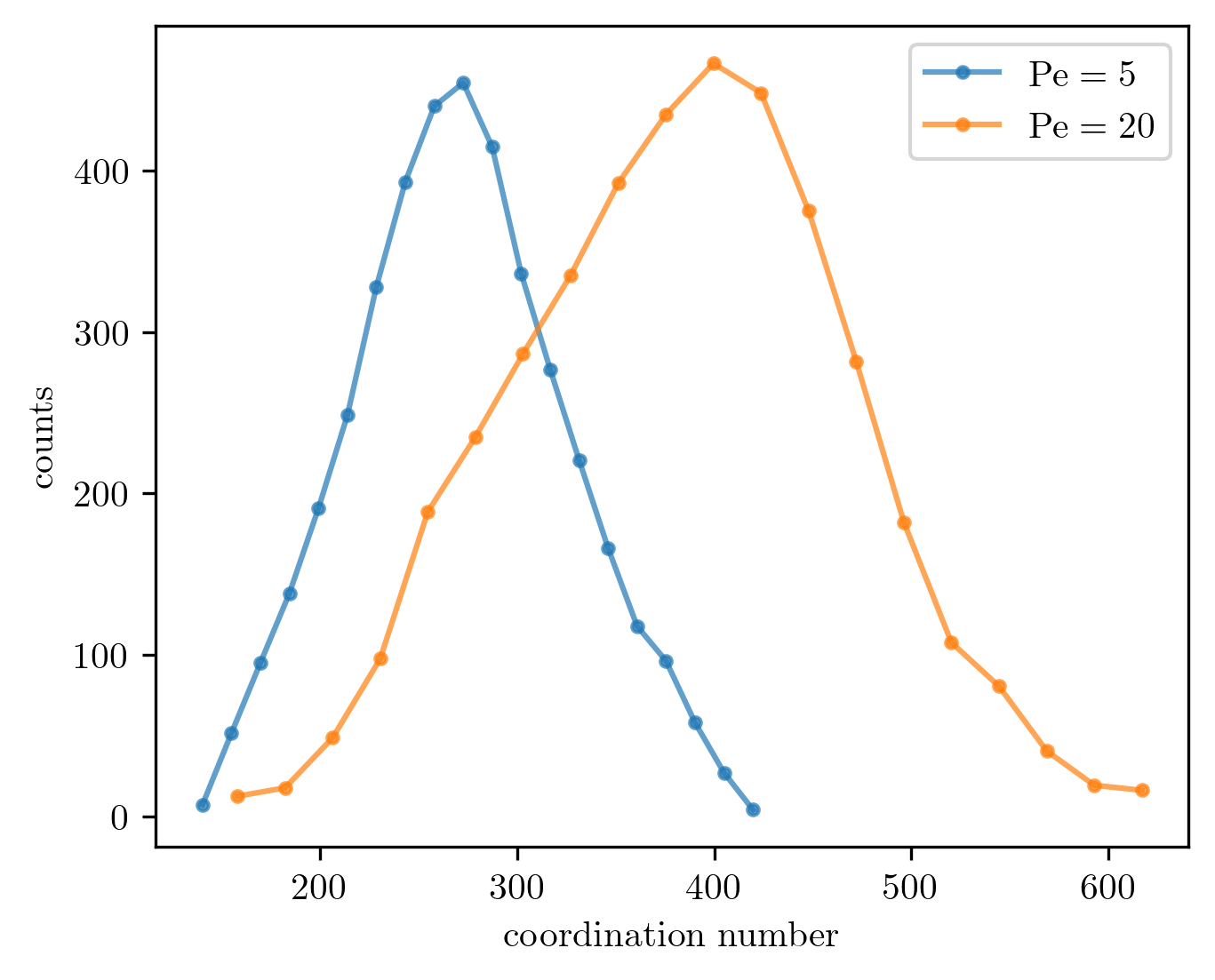}
    \caption{ {Distribution of the number of particles (coordination number) within 
    a sphere of fixed radius of $5\,\sigma$ in a system with $\phi=0.25$ and 
    $\epsilon=0.3$ and two Pe values. }
    }
    \label{fig:coord-dist}
\end{figure}

 {
In the main text we mentioned that  the gel gets thicker at higher Pe. 
The reason for this is that  
the activity breaks the metastable gel configurations and facilitates the formation of more organized local crystal structures, which result in thicker branches. In a first step, we observed this behaviour from 
the direct inspection of the dumbbell configurations and evolution. 
In a second step, we quantified how the thickness of the structures evolves 
by computing the distributions of beads inside a sphere of radius $r=5\sigma$ centered at 
each bead. We report the result at two different Pe and same density $\phi=0.25$ in the gel phase in Fig.~\ref{fig:coord-dist}. 
As expected, the distribution mean shifts toward higher coordination numbers with increasing activity, indicating that the structures are indeed thicker. 
}

\section{Phase Diagrams at varying attraction strength}

\begin{figure}[h!]
    \centering
    \includegraphics[width=0.8\linewidth]{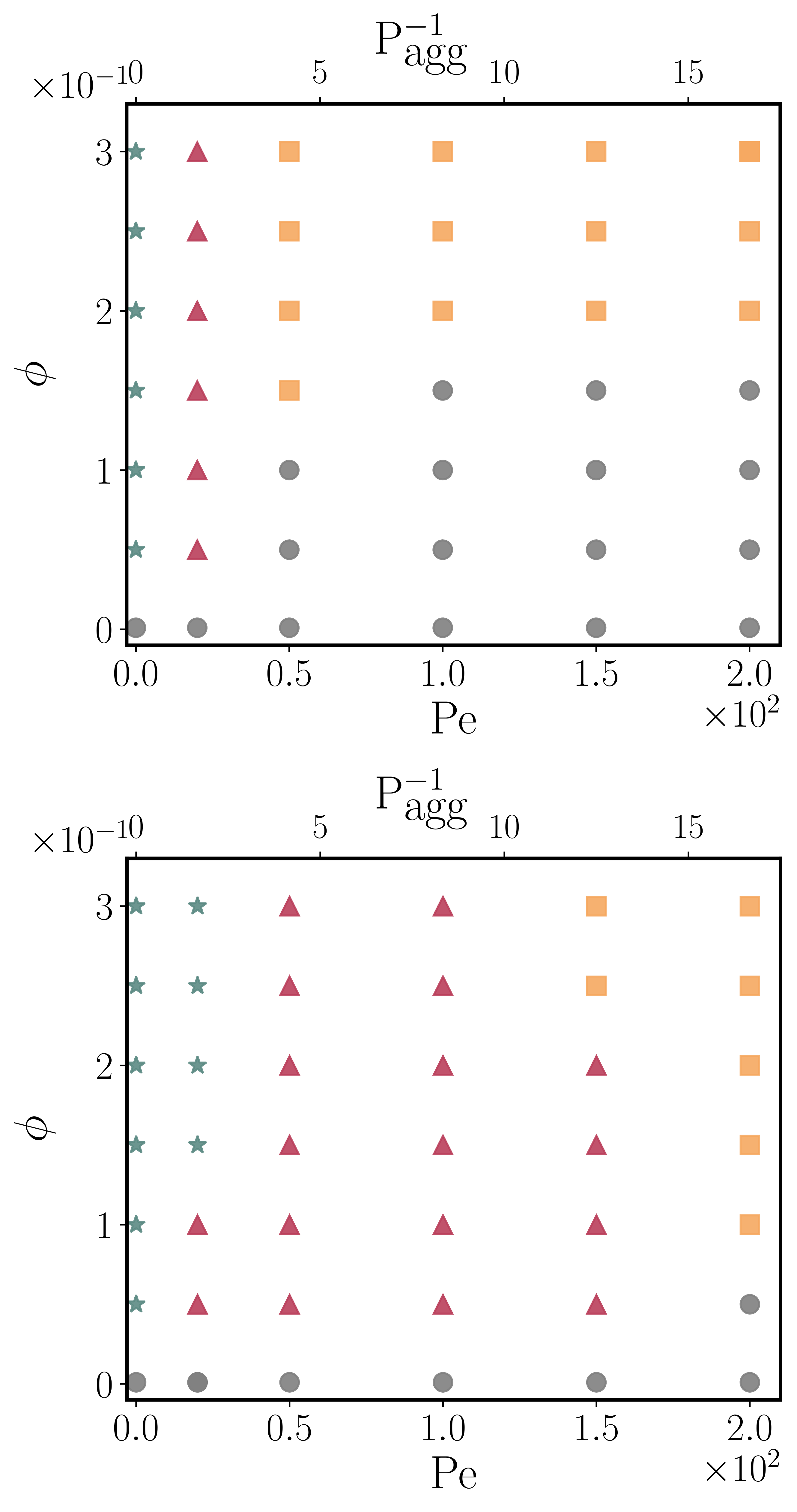}
    \caption{ {{\bf Phase diagrams at different attraction strengths}, $\epsilon=0.2$ (above)
    and $\epsilon=0.2$ (below). The symbols and color codes are the same as the ones in Fig.~\ref{fig:4}, 
    with green stars representing the gel, red triangles the phase separated region, gray bullets the disordered phase and yellow squares the percolating network. }}
    
    \label{fig:appendix1}
\end{figure}

 {In this Appendix we present two other phase diagrams,  in the plane Pe - $\phi$, for $\epsilon=0.2$ (panel above in Fig.~\ref{fig:appendix1}) and $\epsilon=0.4$ (panel below in the same figure), to be compared to the case $\epsilon=0.3$ shown in Fig.~\ref{fig:4} in the body of the paper. The horizontal axes are also parametrized by $P_{\rm agg}^{-1}$ following the scale at the top of the plots. Naturally, the attraction stabilizes the 
gel (green stars) and phase separated (red triangles) phases, 
which extend towards larger values of Pe, at fixed $\phi$, for increasing~$\epsilon$.}

\section{The cluster's trajectory}
\label{app:trajectory}

\begin{figure}
        \centering
        \includegraphics[width=0.8\linewidth]{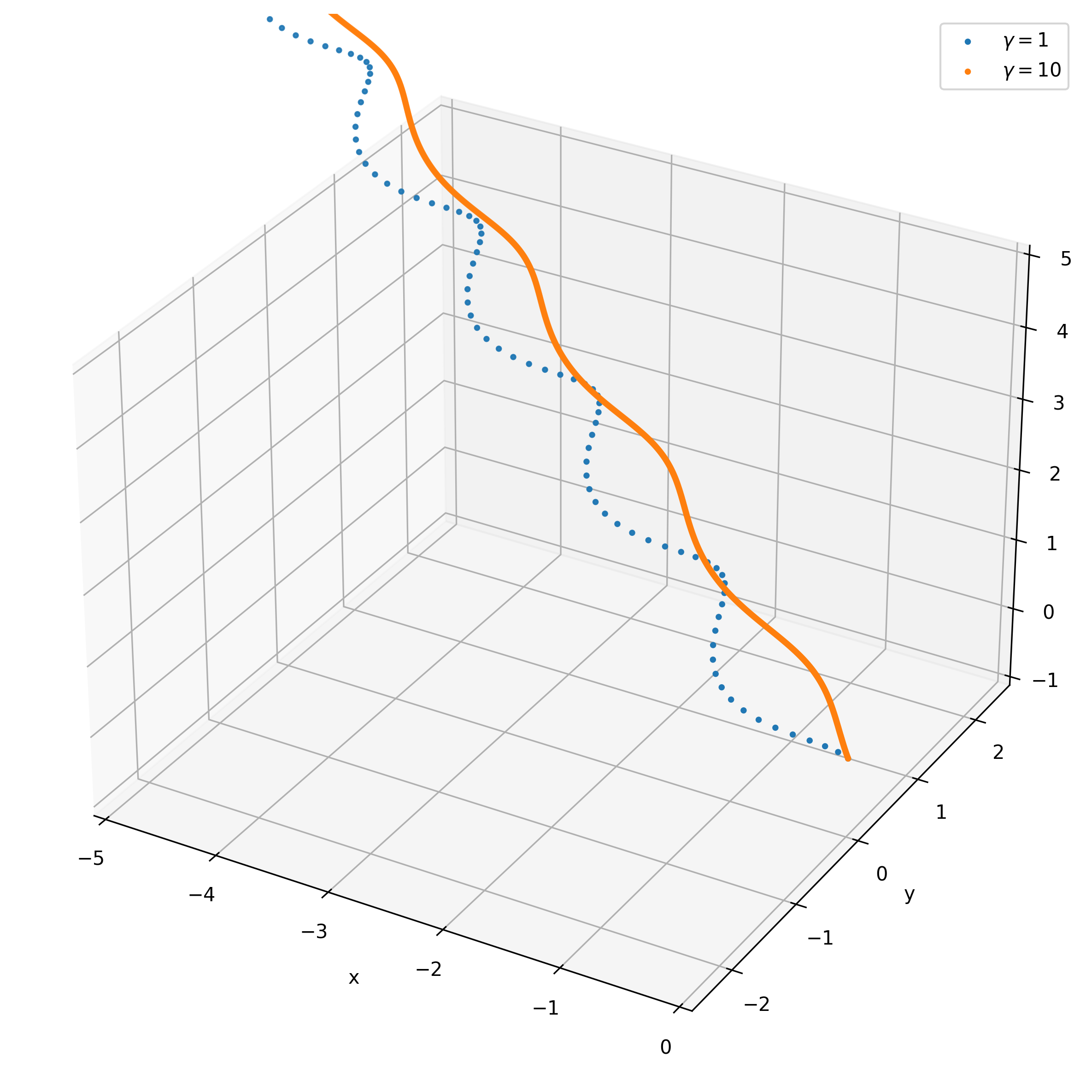}
        \caption{ {Trajectory of a single cluster with $N=3278$ beads and using different values of the damping coefficient $\gamma$, reported in the legend.}}
        \label{fig:trj-diff-gamma}
    \end{figure}

 {
In Fig.~\ref{fig:trj-diff-gamma} we plot the trajectory of a selected cluster using molecular dynamics with different damping coefficient $\gamma$. Not surprisingly, the weaker the damping the larger the oscillations in the motion around the propulsion axis, 
called $\hat{\bm{k}}$ in the analysis of Sec.~\ref{sec:single-cluster}.
}

\clearpage

\bibliographystyle{unsrt}
\bibliography{biblio}
\end{document}